\documentclass[final,5p,times,twocolumn]{elsarticle}

\usepackage{amssymb}
\usepackage[numbers]{natbib}
\usepackage[most]{tcolorbox}
\usepackage{dirtytalk}
\usepackage{fontawesome}
\usepackage{pifont}
\usepackage{balance}
\usepackage[T1]{fontenc}
\usepackage{ragged2e}
\usepackage{makeidx}
\usepackage{multicol}
\usepackage{multirow}
\usepackage{lipsum}
\usepackage{footnote}

\usepackage{tabularx}
\usepackage{colortbl}
\usepackage{hyperref}
\usepackage{footmisc}
\usepackage{amsthm}

\definecolor{anti-flashwhite}{rgb}{0.95, 0.95, 0.96}
\definecolor{beige}{rgb}{0.96, 0.96, 0.86}
\definecolor{floralwhite}{rgb}{1.0, 0.98, 0.94}
\definecolor{gainsboro}{rgb}{0.86, 0.86, 0.86}
\definecolor{ghostwhite}{rgb}{0.97, 0.97, 1.0}
\definecolor{honeydew}{rgb}{0.94, 1.0, 0.94}
\definecolor{isabelline}{rgb}{0.96, 0.94, 0.93}
\definecolor{ivory}{rgb}{1.0, 1.0, 0.94}
\definecolor{magnolia}{rgb}{0.97, 0.96, 1.0}
\definecolor{mintcream}{rgb}{0.96, 1.0, 0.98}
\definecolor{pearl}{rgb}{0.94, 0.92, 0.84}
\definecolor{whitesmoke}{rgb}{0.90, 0.90, 0.90}

\def\tsc#1{\csdef{#1}{\textsc{\lowercase{#1}}\xspace}}
\tsc{WGM}
\tsc{QE}
\tsc{EP}
\tsc{PMS}
\tsc{BEC}
\tsc{DE}

\journal{Accepted to appear in Journal of Systems and Software, 2021.}

\begin{document}

\begin{frontmatter}



\let\WriteBookmarks\relax
\def\floatpagepagefraction{1}
\def\textpagefraction{.001}

\title{Automated Identification of Security Discussions in Microservices Systems: Industrial Surveys and Experiments}






\author[1]{Ali Rezaei-Nasab}
\ead{rezaei.ali.nasab@gmail.com}
\address[1]{School of Computer Science, Wuhan University, 430072 Wuhan, China}

\author[2]{Mojtaba Shahin\corref{cor1}}
\ead{mojtaba.shahin@monash.edu}
\address[2]{Department of Software Systems and Cybersecurity, Faculty of IT, Monash University, 3800 Melbourne, Australia}

\author[1]{Peng Liang\corref{cor1}}
\ead{liangp@whu.edu.cn}

\author[3]{{Mohammad Ehsan} Basiri}
\ead{basiri@sku.ac.ir}
\address[3]{Department of Computer Engineering, Shahrekord University, 64165478 Shahrekord, Iran}

\author[1]{{Seyed Ali} Hoseyni Raviz}
\ead{s.ali.hoseyni@gmail.com}

\author[2]{\\Hourieh Khalajzadeh}
\ead{hourieh.khalajzadeh@monash.edu}

\author[1]{Muhammad Waseem}
\ead{m.waseem@whu.edu.cn}

\author[4]{Amineh Naseri}
\ead{naseri.amine@sirjantech.ac.ir}
\address[4]{Department of Computer Engineering, Sirjan University of Technology, 7813733385 Sirjan, Iran}

\cortext[cor1]{Corresponding authors}

\begin{abstract}
Lack of awareness and knowledge of microservices-specific security challenges and solutions often leads to ill-informed security decisions in microservices system development. We claim that identifying and leveraging security discussions scattered in existing microservices systems can partially close this gap. We define security discussion as \say{\textit{a paragraph from developer discussions that includes design decisions, challenges, or solutions relating to security}}. We first surveyed 67 practitioners and found that securing microservices systems is a unique challenge and that having access to security discussions is useful for making security decisions. The survey also confirms the usefulness of potential tools that can automatically identify such security discussions. We developed fifteen machine/deep learning models to automatically identify security discussions. We applied these models on a manually constructed dataset consisting of 4,813 security discussions and 12,464 non-security discussions. We found that all the models can effectively identify security discussions: an average precision of 84.86\%, recall of 72.80\%, F1-score of 77.89\%, AUC of 83.75\% and G-mean 82.77\%. DeepM1, a deep learning model, performs the best, achieving above 84\% in all metrics and significantly outperforms three baselines. Finally, the practitioners' feedback collected from a validation survey reveals that security discussions identified by DeepM1 have promising applications in practice.
\end{abstract}




\end{frontmatter}



\section{\textbf{Introduction}}  \label{introduction} 
The Microservices Architecture (MSA) style, as one of the latest trends in software design, aims to develop scalable software systems as a collection of small services (i.e., microservices) that can be deployed independently \cite{dragoni2017microservices}, \cite{fowler2014microservices}. Microservices are open to a diverse range of interpretations, and little consensus exists regarding their characteristics (e.g., the size of microservices) \cite{jamshidi2018microservices}. However, microservices can be generally characterized by providing a limited amount of functionality, communicating with each other via light-weight messaging approaches (e.g., HTTP APIs), being built around autonomous business capabilities, and being maintained and tested in high isolation \cite{dragoni2017microservices}, \cite{fowler2014microservices}, \cite{newman2015building}. Faster deployment, improved scalability, and greater autonomy are three main benefits of microservices systems (the systems that adopt the MSA style) \cite{jamshidi2018microservices}. 

Over the past few years, there has been ongoing research on different aspects of MSA. A significant body of literature reports migration techniques to move from a monolith to an MSA (e.g., \cite{balalaie2016microservices}, \cite{Taibi2017}) \cite{di2019architecting}. Others investigate testing and monitoring strategies for microservices systems (e.g., \cite{heorhiadi2016gremlin}, \cite{heinrich2017performance}). Another line of research focuses on {the required changes} in organizational culture and structure to adopt the MSA style (e.g., \cite{jamshidi2018microservices}, \cite{haselbock2018expert}). 

However, security in MSA remains an open issue and has received insufficient research attention \cite{dragoni2017microservices}, \cite{di2019architecting}, \cite{waseem2020systematic}. 
The security issues associated with the MSA style are numerous, which could not be found or are of much less interest in traditional monoliths and service-oriented architectures \cite{sun2015security}, \cite{yarygina2018overcoming}, \cite{yu2019survey}, \cite{esposito2016challenges}. They range from the need for establishing trust between individual microservices because a single compromised service may maliciously impact the entire system, to the increasing chance of attack surfaces as microservices can be developed and deployed by different technologies and tools (e.g., containers), to the greater difficulty to guarantee the security of tens or even hundreds of microservices running in {operation}. Furthermore, as the MSA style is a new paradigm and still evolving, there is a knowledge gap among practitioners and organizations on designing and implementing a secure microservices system \cite{ghofrani2018challenges}, \cite{zimmermann2017microservices}, \cite{Pereira2019SecMec}, \cite{nadareishvili2016microservice}. More significant unknowns exist around (new) technologies and tools (e.g., containers) that are used to develop and deploy microservices \cite{McAfeeSecReport2018}, \cite{scott2018MSACont}, \cite{zhang2019microservice}. This is because a lot of confusion exists in the industry {about the potential security capabilities} of those technologies and tools and the best practices to configure them for security purposes and address their security vulnerabilities \cite{bogner2019microservices}, \cite{knoche2019drivers}, \cite{bavskarada2018architecting}.

Lack of awareness and knowledge of security in microservices systems often leads to ill-informed security decisions in microservices system development. This can become a real issue in the software industry because a growing number of organizations have employed or plan to use this architectural style to modernize their software solutions or develop new applications \cite{di2019architecting}. Our position to (partially) address this gap is to automatically identify and leverage security discussions from existing microservices systems. We define \textbf{security discussion} as \say{\textit{a paragraph from developer discussions (conversations) that includes design decisions, challenges, or solutions relating to security}}. It is argued that accessing and learning past design discussions and decisions (e.g., security decisions) can help make better design decisions (e.g., \cite{zimmermann2017microservices}, \cite{Pereira2019SecMec}, \cite{capilla201610}, \cite{viviani2019locating}). However, we believe that security discussions collected from non-microservices systems may not be useful and practical to increase awareness and knowledge of security in microservices systems among microservices practitioners. To prove our claim, we first need to ensure that securing microservices systems is a unique challenge. This shows that security discussions that occur between microservices developers are different from other types of software systems. We then need to confirm if such security discussions will be useful in securing microservices systems (e.g., raising practitioners' awareness of microservices-specific security solutions).


{Such security discussions can be found in the issue tracking systems of commercial and open-source microservices systems.} Stack Overflow posts can be another source to identify security discussions on microservices systems. Research shows that such artifacts include a wide range of information, such as user stories, requirements, design decisions, design solutions, design rationale, bug reports, etc (e.g., \cite{bhat2017automatic}, \cite{viviani2019locating}, \cite{tsay2014let}, \cite{lopez2019anatomy}). {Hence, issue tracking systems and Stack Overflow posts are ideal places for microservices practitioners to share and communicate the knowledge about security issues, practices, design decisions, and justify their security decisions \cite{palacio2019learning}, \cite{lopez2019anatomy}.} {However, manually identifying security discussions is challenging for practitioners because it is error-prone, takes significant time for practitioners, and requires substantial domain expertise \cite{bettaieb2019decision}, \cite{li2020ontology}. Hence, there is a need for an automated approach to identify security contents in such data sources.}

In this paper, we first identified ten open-source projects from GitHub designed based on the MSA style. We then conducted an online survey completed by 67 microservices practitioners to solicit their perspectives on security in microservices systems. The survey participants confirmed that securing microservices systems is a unique challenge for them. {They also found that previous security discussions collected from existing microservices systems are useful for making security decisions.} Based on the insights from the survey, we developed twelve Machine Learning (ML) and three Deep Learning (DL) models to automatically identify security discussions from microservices developer discussions ({GitHub} issues and Stack Overflow posts). 
To evaluate the performance of the ML and DL models (henceforth learning models), we manually built and labeled a dataset of 17,277 paragraphs. The 17,277 paragraphs, including 4,813 security discussions and 12,464 non-security discussions, were collected from two data sources: 1,692 issue discussions from five open-source projects that adopt the MSA style, and 498 Stack Overflow posts with \say{microservices} and \say{security} tags. The experimental results show that {all learning models} are promising, achieving, on average, a precision of 84.86$\%$, recall of 72.80$\%$, F1-score of 77.89$\%$, AUC (Area Under the Receiver Operating Characteristic Curve) of 83.75$\%$, and G-mean 82.77$\%$. Of the 15 learning models, DeepM1, a DL model, performs the best, achieving 86.73$\%$ precision, 84.25$\%$ recall, 85.47$\%$ F1-score, 89.63$\%$ AUC, and 89.07$\%$ G-mean. DeepM1 also outperforms three state-of-the-art baselines with improvements ranging from 1.018x to 3.756x in all metrics.

We conducted another survey (validation survey) to show the usefulness and actionability of the results produced by DeepM1 in practice. Generally, 68\%-78\% of the validation survey respondents perceived that the security discussions detected by DeepM1 could have seven practical applications in the design and development of secure microservices systems. The notable applications are: the detected security discussions can help them make informed security decisions in the future or refine the existing sub-optimum security decisions and provide hints/clues to locate critical issues (e.g., security bugs, security mistakes, etc.) faster in microservices systems.

The key contributions of this paper are that: \textbf{(1)} We gain a better understanding of security concerns in microservices systems. \textbf{(2)} This is the first work that develops ML and DL models to discriminate security discussions from non-security discussions in developer discussions of microservices systems. \textbf{(3)} We construct a dataset consisting of 4,813 microservices security discussions and 12,464 non-security discussions. \textbf{(4)} We freely release the implementation of the models and the datasets used in this paper online \cite{onlinedataset}.

\textbf{Paper Organization}: In Section \ref{sec:motivatinsc}, we provide a motivating scenario. Section \ref{securitydiscussions} describes the process of locating MSA-based projects from GitHub and building a security discussions dataset. Section \ref{secresearch} describes our methodology, including survey and experiments. Section \ref{secfindings} reports the findings. Section \ref{secDiscuss} reflects on the findings, and Section \ref{secthreats} reports the threats to validity. We examine the related work in Section \ref{secRelatework}. Section \ref{secConclusion} concludes the paper and outlines some future works.

\begin{figure*}[h]
\centering
		\includegraphics[scale=.55]{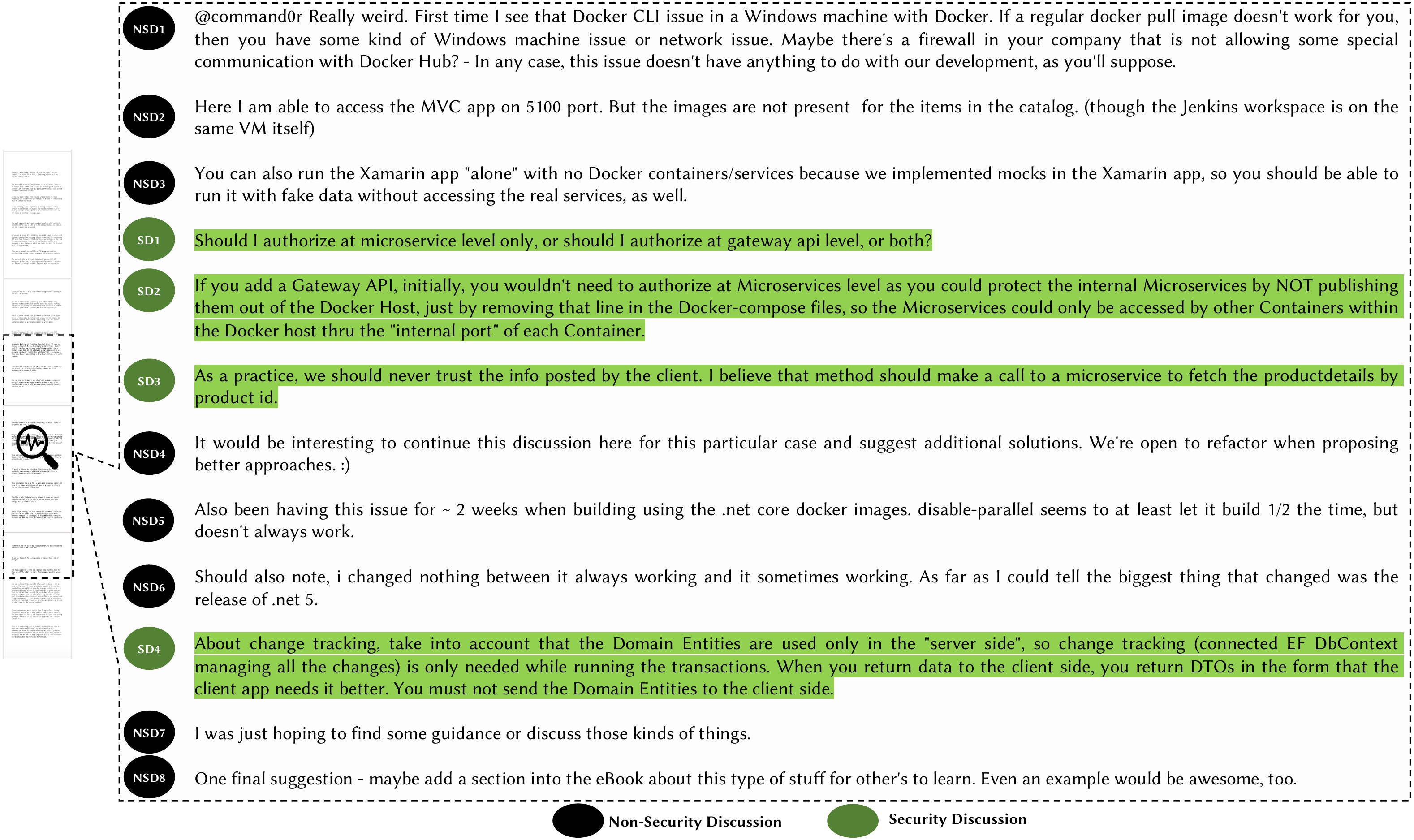}
	\caption{An excerpt of developer discussions in the eShopOnContainers project}
	\label{FIG:motivatingscenario}
\end{figure*}

\section{Motivating Scenario}\label{sec:motivatinsc}
Bob is working for a company that develops a wide range of software systems in different domains. A few months ago, Bob was involved as a senior software engineer in developing an e-Shop web application (called the eShopOnContainers project \footnote{\href{https://bit.ly/3uRN2qu}{\url{https://bit.ly/3uRN2qu}}}) in the company, which was designed based on the MSA style. Recently, the company won a tender to migrate a large-scale monolithic system to the MSA style (we call it Project C). Given that Bob was involved in the eShopOnContainers project, the company has assigned Bob as a team leader of Project C. Bob's observations from the eShopOnContainers project are:
\begin{enumerate}
   \item Conversations between developers of the eShopOnContainers project captured in the issue tracking system include various security information. An excerpt of developer discussions in the eShopOnContainers project is shown in Figure \ref{FIG:motivatingscenario}.
    \item Some of the security challenges and concerns in the MSA style are unique, which have not been faced before by developers of the eShopOnContainers project. Therefore, the strategies adopted to address those challenges were new.
    \item Although some security issues faced by developers of the eShopOnContainers project were not new and had common security solutions, the developers repeatedly had those security issues and mistakes \cite{tahaei2021security}. The developers frequently looked at their previous discussions to recognize which ones were about security and understand how those security issues and mistakes were solved.
\end{enumerate}

Inspired by the idea of the Open Web Application Security Project (OWASP)\footnote{\href{https://owasp.org/}{\url{https://owasp.org/}}} and his observations from the eShopOnContainers project, Bob thinks that the security of Project B can be improved if its developers' awareness is increased about the most critical security issues in microservices systems, their impacts, and their corresponding solutions \cite{tahaei2021security}. Further to this, such security information can be imported to (security) knowledge management tools used in the company \cite{capilla201610}. In such a scenario, knowing where are the security information is a direct prerequisite for organizing it to a structured or company-specific format \cite{abualhaija2019machine}.

In the light of the above scenario, Bob asks a member of Project C, Alex, a newcomer to the company, to collect security discussions from developer discussions in the eShopOnContainers project. Given Alex has limited experience in (security in) microservices, he may face the following challenges in this task. (1) He needs the domain knowledge to distinguish developer security discussions from non-security discussions in microservices systems \cite{taibi2017microservices}. Some security discussions (e.g., SD4 in Figure \ref{FIG:motivatingscenario}) do not have (common) security-related terms. Some non-security discussions (e.g., NSD1 in Figure \ref{FIG:motivatingscenario}) include common security-related terms. (2) Doing this task for Alex is both time-consuming and tedious. For example, a developer discussion \footnote{\href{https://github.com/dotnet-architecture/eShopOnContainers/issues/107}{\url{https://bit.ly/2FXeZdg}}\label{eshop107}}, which includes 293 comments, may take 10 hours for Alex to deeply read the entire of that developer discussion \cite{viviani2019locating}.

\begin{figure*}[t]
	\centering
		\includegraphics[scale=.80]{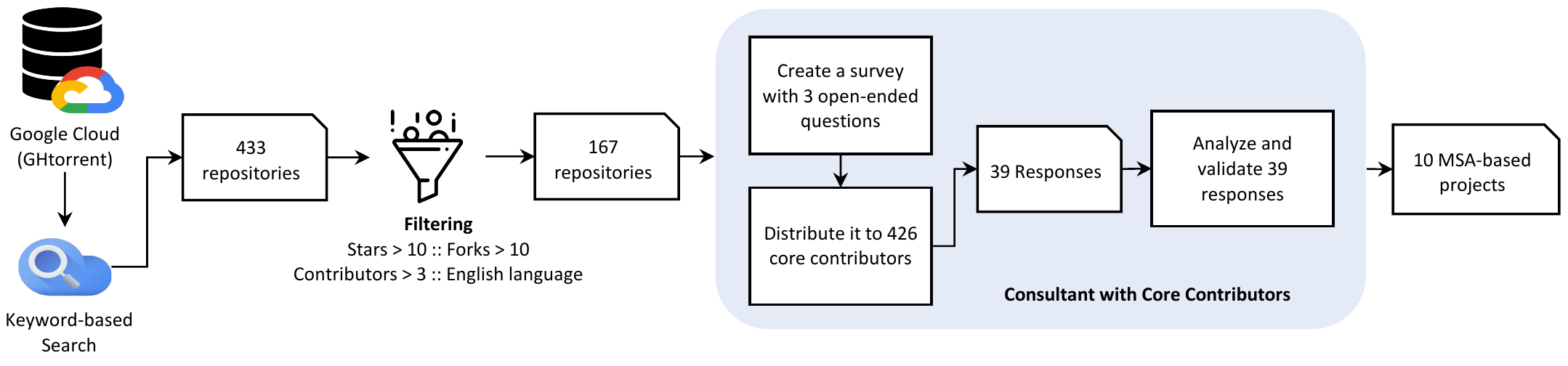}
	\caption{Process of locating 10 open-source projects that adopt the MSA style}
	\label{FIG:locatingMSAprojects}
\end{figure*}

Hence, a tool that can automatically and accurately identify security discussions, including critical security issues in microservices systems, their impacts, and corresponding solutions, can help develop secure microservices systems.The output of such a tool has educational benefits for microservices practitioners and raise their awareness of the different aspects of security in microservices systems.

\section{A Security Discussions Dataset}\label{securitydiscussions}
Our goal is to (partially) bridge the knowledge gap among practitioners in securing microservices systems by automatically identifying security discussions from previous microservices developer discussions. To this end, we need to create a dataset of security discussions. As discussed in the Introduction section, security discussions can be found in the issue tracking systems of commercial and open-source systems and Stack Overflow posts. Security discussions in commercial software systems usually are not accessible to the public. Hence, we decided to build our dataset based on security discussions in open-source microservices systems on GitHub and Stack Overflow. As there is no reliable information indicating which open-source systems on GitHub are designed based on the MSA style, we first had to identify such systems on GitHub. Section \ref{findmsaproject} describes the process of finding microservices systems on GitHub. Section \ref{secBuilddataset} reports how a security discussions dataset is created from developer discussions in five microservices systems on GitHub and Stack Overflow posts.

\subsection{\textbf{Phase I: Locate Microservices Systems in GitHub}}\label{findmsaproject}

The process of finding microservices systems {is composed of} two steps: keyword-based search and consulting with core contributors.

\subsubsection{\textbf{Keyword-based Search}}\label{seckeywordsearch}

Figure \ref{FIG:locatingMSAprojects}\ shows the process of {locating} MSA-based projects. As shown in Figure \ref{FIG:locatingMSAprojects}\footnote{Some of the icons are from \href{https://bit.ly/3dta9iM}{\url{https://bit.ly/3dta9iM}}}, we first executed a search string on the GHTorrent dump hosted on Google Cloud version 1/4/2018 \cite{gousios2017mining}. Specifically, we searched the following search string on the title and description of open-source projects.
The search process resulted in 2,690 repositories. Previous research showed that the number of forks and stars could be a good indicator of popular repositories \cite{joshi2019rapidrelease}. We then used these factors to obtain the repositories that {had} such properties. We chose the repositories whose fork and star were more than 10, resulting in 433 repositories.

\begin{center}
\fbox{\begin{minipage}{20.5em}
{\small \centering
\say{microservice} OR \say{micro service} OR \say{micro-service}\\OR\\\say{Microservice} OR  \say{Micro service} OR \say{Micro-service}\\}
\end{minipage}}
\end{center}

\subsubsection{\textbf{Consult with Core Contributors}}\label{secconsult} 
Having \say{microservices} term in the title or description of a project does not guarantee that the project is designed based on the MSA style. Hence, we needed to make sure if those 433 projects follow the MSA style. We only focused on the repositories with more than three contributors and with the English language. This decreased the number of repositories from 433 to 167. Afterward, we tried to communicate with the core contributors of the 167 projects and directly asked them whether or not these projects follow the MSA style. We define the core contributors of a project as the top three contributors who have the most commits in the project, and their email addresses are publicly available. We supposed that this type of contributor has a broad view of a project and is more likely to know its design and structure. Then, we checked their GitHub profiles or searched their names to find their publicly available emails. We finally got the emails of 426 core contributors from the 167 projects, and we emailed and asked them to answer the following questions: 

\begin{figure*}[h]
\centering
\includegraphics[scale=0.85]{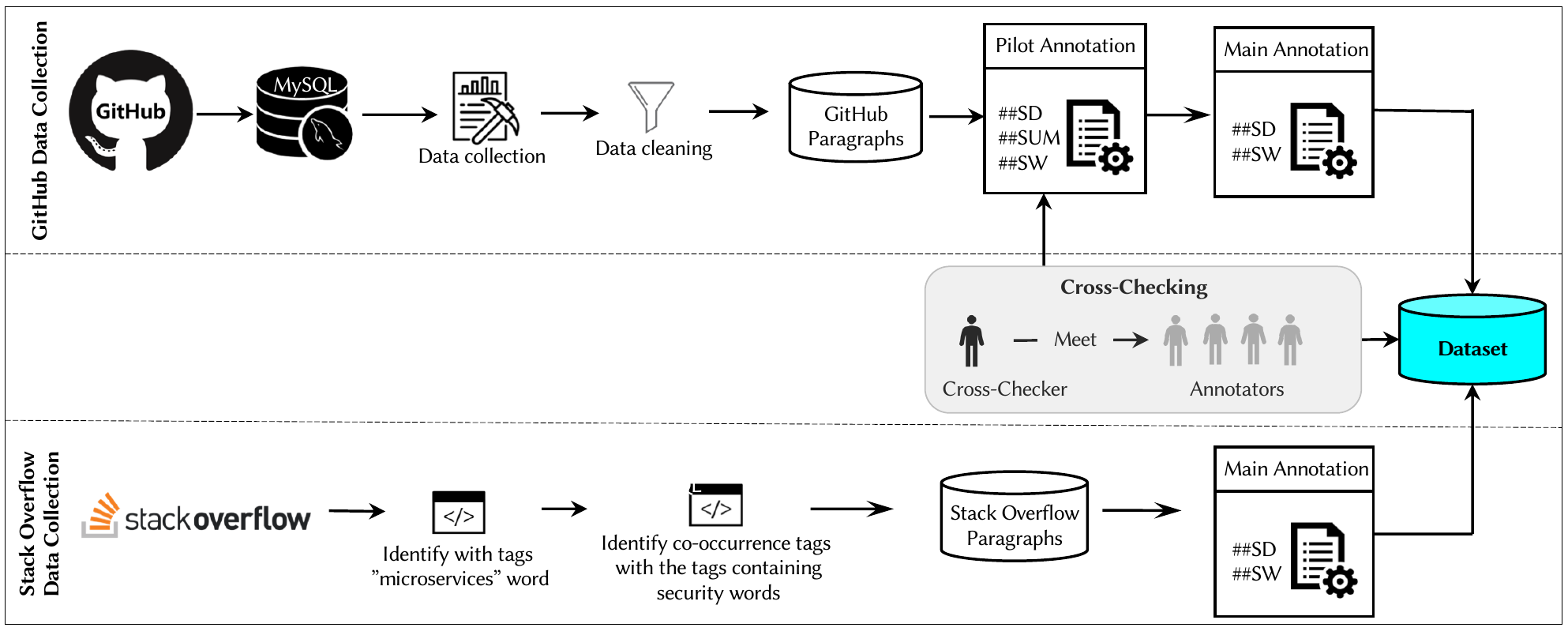}
\caption{Process of building a Security Discussions Dataset}
\label{FIG:buildDS}
\end{figure*}

\begin{enumerate}
\item \textit{Can you please confirm if our interpretation is correct that this project (URL of the project) is designed by following the MSA style? Note that if the projects (e.g., frameworks or tools) support the development of microservices systems but are not designed based on the MSA style, please clarify it.}

\item \textit{If our interpretation is correct, what features or characteristics of the architecture of this open-source project show that the MSA style has been used?}

\item \textit{Optional question: Do you know any other open-source projects that are designed by following the MSA style?}
\end {enumerate}

Out of the 426 core contributors, 39 replied to our emails (response rate: 9\%). Table \ref{tablecorecontributor} shows two core contributors’ responses. It should be noted that while some respondents answered \say{\textbf{Yes}} to question \textbf{(1)} and claimed that the asked project follows the MSA style, they did not provide a detailed answer or any answer to question \textbf{(2)}. To avoid any misinterpretation and possible risks, we did not consider this type of projects as a microservices project. The analysis of core contributors’ responses revealed that 10 open-source projects had been designed by following the MSA style. The basic information about these projects is shown in Table ~\ref{tab:microservices projects}.

\subsection{\textbf{Phase II: Construct a Security Discussions Dataset}}\label{secBuilddataset}
This section describes the construction of a dataset {consisting of 17,277 labeled paragraphs from two sources: developer discussions of five open-source microservices systems on GitHub and Stack Overflow posts. This dataset is later in Section \ref{secexp} used for training and testing learning models. Figure \ref{FIG:buildDS} shows the process of constructing this dataset.}

\begin{table*}[h]
\centering
\caption{Core contributors’ responses from eShopOnContainers and deep-framework}\label{tablecorecontributor}
{\small
\begin{tabular}{|c|p{16.2cm}|}
\hline 
            \multicolumn{1}{|c}{\textbf{}}       & \multicolumn{1}{c|}{\textbf{Answers}} \\ \hline \hline
{\hfil \multirow{3}{*}{{\rotatebox[origin=c]{90}{\parbox[c]{3cm}{\centering  {\footnotesize eShopOnContainers project}}}}}} & \textbf{Answer to Question (1):} That’s right, that project is based on microservices architecture. In fact, [there is] a whole book about microservices architecture with Microsoft .NET where eShopOnContainers is the reference application. Related free Microsoft eBook: \href{https://aka.ms/microservicesebook}{{https://bit.ly/3aF04Od}}. Other Microsoft books/guidance: \href{https://dotnet.microsoft.com/learn/dotnet/architecture-guides}{{https://bit.ly/2Ku89KH}}. \\ \cline{2-2} 
                  &  \textbf{Answer to Question (2):} (a) Microservice’s autonomy, including an autonomous database per microservice. (b) Use of API Gateways. (c) Use of Container Orchestrator when deploying to production (Kubernetes). (d) Asynchronous communication between updates/transactional operations between microservices to achieve eventual consistency between their data. You can see many more points in the Book, since it is related to this reference application. \\ \cline{2-2} 
                  & \textbf{Answer to Question (3):} This one by Microsoft as well (I was not involved in this one, though): \href{https://github.com/mspnp/microservices-reference-implementation}{{https://bit.ly/2Vw62wo}} \\ \hline
\multirow{3}{*}{{\rotatebox[origin=c]{90}{\parbox[c]{2.5cm}{\centering {\footnotesize deep-framework project}}}}} & \textbf{Answer to Question (1):}
Yes, deep-framework is designed using microservices architectural patterns. To be clear, this is an abstracted set of javascript / nodejs libraries that simplify development and provisioning of cloud resources (AWS to be specific). \\ \cline{2-2} 
                  & \textbf{Answer to Question (2):}
Componentization via Services, Products not Projects, Decentralized Governance, Decentralized Data Management, Infrastructure Automation and more. We have been using Micro Frontends since 2015, way before Cam Jackson published his article from 2019. \\ \cline{2-2} 
                  & \textbf{Answer to Question (3):}
Yeah, we can proudly point to another repository that uses microservices architectural patterns called AWS Landing Zone as Terraform Module: \href{https://github.com/terrahubcorp/terraform-aws-landing-zone}{{https://bit.ly/3iVXh7Q}}  \\ \hline 
\end{tabular}}
\end{table*}

\begin{center}
\begin{table*}[h]
\centering
\caption{List of identified microservices systems on GitHub. Number of Issues (\faExclamationCircle); Number of Releases (\faTags); Number of Contributors (\faGroup); Number of Stars (\faStarO); Number of Forks (\faCodeFork); Active Date (\faCalendarPlusO); Line of Codes (\faCode); Languages (\faLanguage)}
 \label{tab:microservices projects}
{\scriptsize
\renewcommand{\arraystretch}{1.2}
\begin{tabular}{ccccccccc}
\hline 
\multicolumn{1}{c}{\begin{tabular}[c]{@{}c@{}}\textbf{Project Name} \\ (URL) \end{tabular} }                                                                                                                                              & \multicolumn{1}{c}{\normalsize \faExclamationCircle} & \multicolumn{1}{c}{\normalsize \faTags} & \multicolumn{1}{c}{\normalsize \faGroup} & \multicolumn{1}{c}{\normalsize \faStarO} & \multicolumn{1}{c}{\normalsize \faCodeFork} & \multicolumn{1}{c}{\normalsize \faCalendarPlusO} & \multicolumn{1}{c}{\normalsize \faCode}  & \multicolumn{1}{c}{\normalsize \faLanguage}                   \\ \hline 
\begin{tabular}[c]{@{}c@{}}\textbf{goa}     \\ \href{https://bit.ly/2Vz0GjV}{{(https://bit.ly/2Vz0GjV)}}\end{tabular}                         & 2442                            & 31                              & 73                               & 4.1k                              & 450                               &    2014-now          & 82,193            & GO                          \\ 
\begin{tabular}[c]{@{}c@{}}\textbf{eShopOnContainers }          \\ \href{https://bit.ly/3eMUFYy}{{(https://bit.ly/3eMUFYy)}}\end{tabular}     & 1200                            & 15                              & 98                               & 16.2k                             & 6.8k                              &       2016-now       & 136,963           & C\#, Javascript, HTML       \\ 
\begin{tabular}[c]{@{}c@{}}\textbf{microservices-demo }          \\ \href{https://bit.ly/3cKLxln}{{(https://bit.ly/3cKLxln)}}\end{tabular}    & 805                             & 13                              & 43                               & 2.6k                              & 2.6k                              &      2016-now        & 18,828            & Shell, Python, HCL, Ruby    \\ 
\begin{tabular}[c]{@{}c@{}}\textbf{scalecube-services}            \\ \href{https://bit.ly/3bBBWgq}{{(https://bit.ly/3bBBWgq)}}\end{tabular}   & 689                             & 96                              & 18                               & 477                               & 72                                &      2015-now        & 11,764            & Java                        \\ 
\begin{tabular}[c]{@{}c@{}}\textbf{moleculer}                    \\ \href{https://bit.ly/3bytGxK}{{(https://bit.ly/3bytGxK)}}\end{tabular}    & 649                             & 87                              & 65                               & 4k                                & 383                               &      2017-now        & 93,242            & Javascript                  \\ 
\begin{tabular}[c]{@{}c@{}}\textbf{deep-framework}              \\ \href{https://bit.ly/3cO6o79}{{(https://bit.ly/3cO6o79)}}\end{tabular}     & 640                             & 22                              & 9                                & 532                               & 75                                &       2015-now       & 920,806           & HTML, Javascript, CSS \\ 
\begin{tabular}[c]{@{}c@{}}\textbf{light-4j }                    \\ \href{https://bit.ly/2Y3eEwe}{{(https://bit.ly/2Y3eEwe)}}\end{tabular}    & 637                             & 104                             & 25                               & 3k                                & 496                               &      2016-now        & 50,099            & Java, Objective-J           \\ 
\begin{tabular}[c]{@{}c@{}}\textbf{apollo}                   \\ \href{https://bit.ly/3axm4KM}{{(https://bit.ly/3axm4KM)}}\end{tabular}        & 298                             & 39                              & 37                               & 1.6k                              & 215                               &        2015-now      & 1,828             & Java                        \\
\begin{tabular}[c]{@{}c@{}}\textbf{spring-petclinic-microservices}  \\  \href{https://bit.ly/2S3EXON}{{(https://bit.ly/2S3EXON)}}\end{tabular} & 142                             & 4                               & 24                               & 747                               & 894                               &       2016-now       & 14,506            & Java, HTML, Javascript      \\
\begin{tabular}[c]{@{}c@{}}\textbf{microservice\_workshop}          \\ \href{https://bit.ly/2Y2pNgw}{{(https://bit.ly/2Y2pNgw)}}\end{tabular} & 13                              & 67                              & 4                                & 68                                & 59                                &       2014-now       & 163,237           & Java, C\#, Ruby            \\ \hline 
\end{tabular}
}
\end{table*}
\end{center}

\subsubsection{\textbf{GitHub Data Collection}}\label{secgithubdataset}

\textbf{\textit{Prepare Data}}. We chose five projects in Table ~\ref{tab:microservices projects}, including \textit{goa}, \textit{eShopOnContainers}, \textit{microservices$-$demo}, \textit{light$-$4j}, and \textit{$deep$-$framework$}, as the reference points {for building a security discussions dataset}. These five projects were selected because they are significantly larger than other projects. {Hence, it is highly likely that their contributors had more discussions in issue tracking systems. The decisions and discussions relating to (the design of) a software system can be usually captured in issue tracking systems \cite{viviani2019locating}.} Hence, we leveraged discussions on issue tracking systems. Both open and closed issues were used to increase the chance of finding security discussions. 
We randomly selected 1,692 issues out of 5,724 issues (i.e., confidence level: 95\% and margin of error: 2\% \cite{kadam2010sample}) extracted from the five mentioned projects using the GitHub v3 API on 09/01/2020. Issues may contain different types of information, such as {code snippets} \cite{li2020automatic}. Consequently, {code snippets} were removed from issues. 

We borrowed the idea from Viviani et al. \cite{viviani2019locating} and chose paragraph as the unit of analysis. This can also help avoid possible difficulties (e.g., identifying useful information from lengthy issues) for practitioners \cite{xu2017answerbot}. We divided 1,692 issue discussions into 12,393 paragraphs.

\textbf{\textit{Pilot Study}}. Before annotating the paragraphs extracted from the five mentioned projects, we conducted a pilot study to understand if security discussions can be found in paragraphs. If so, what are the characteristics of security discussions? We randomly selected 575 paragraphs from 200 issues obtained from the 10 projects listed in Table ~\ref{tab:microservices projects}. Then 575 paragraphs were divided between four annotators (i.e., four authors). The annotators were given a coding schema to annotate the paragraphs assigned to them. Another author was responsible for cross-checking the annotated paragraphs. Each paragraph was annotated using the following pattern: If the annotator specified a paragraph as a \say{\textbf{security discussion}}, the annotator had to:
\begin{itemize}
    \item \textit{Write \say{1}, indicating it is a security discussion [\textbf{SD}]}
    \item \textit{Write a summary about the paragraph (one sentence) [\textbf{SUM}]}
    \item \textit{Write the security-related words [\textbf {SW}]}
\end{itemize}
If the given paragraph \textbf{did not} include a \say{security discussion}, the annotators were asked to:
\begin{itemize}
    \item \textit{Write \say{0}, indicating it is a non-security discussion [\textbf{SD}]}
\end{itemize}
 
At the end of the pilot annotation, we found 152 out of the 575 paragraphs as security discussions, which gave us confidence that investigating paragraphs is the right way of identifying security information.

\begin{table*}[h]
\centering
\caption{Six sample paragraphs with their annotations from our dataset including GitHub data and Stack Overflow data.}
	\label{sampleparagraph}
{\small
\begin{tabular}{|c|p{11.8cm}|p{4cm}|}
\hline 
\multicolumn{1}{|l}{\textbf{}}  & \multicolumn{1}{l}{\textbf{Sample Paragraphs}} & \multicolumn{1}{l|}{\textbf{Annotations}} \\ \hline \hline
\multirow{3}{*}{\rotatebox[origin=c]{90}{\parbox[c]{6cm}{\centering GitHub Data}}}               &             {\say{Agree with @mvelosop - Even when the microservices are internal and you could just set the authorization at the API Gateway level, it is a good security practice to set an authorization boundary on each microservice. But it is up to you how you want to balance security vs. simplicity}. [\href{https://github.com/dotnet-architecture/eShopOnContainers/issues/647}{{Taken from issue 647}}\footref{issue647}]}                          & \textbf{Security Discussion?} Yes \newline
\textbf{Security Words:} authorization; security; Gateway                               \\ \cline{2-3} 
                                       &     \say{For OAuth2 Authorization Server, if client application render the SPA to collect user credentials, we need to make sure our Authorization Code service can handle CORS gracefully}. [\href{https://github.com/networknt/light-4j/issues/14}{{Taken from issue 14}}\footref{issue14}]                            &  
                                    \textbf{Security Discussion?} Yes  \newline \textbf{Security Words:} OAuth2; authorization; credential                              \\ \cline{2-3} 
                                       & \say{Until now, I am always thinking to implement websocket, rpc and graphql in the same code based but it look like they are not compatible at all. At this moment, I am seriously thinking to split the current light-java to spin off light-java-rest and then in the future to implement light-java-rpc with websocket support and light-java-graphql. All of them are based on some common libraries within light-java. Nothing has been done yet and I want to hear from you guys}. [\href{https://github.com/networknt/light-4j/issues/8}{{Taken from issue 8}}\footref{issue8}]                                       &             \textbf{Security Discussion?} No                      \\ \hline
\multirow{3}{*}{\rotatebox[origin=c]{90}{\parbox[c]{4cm}{\centering Stack Overflow Data}}}                     & \say{My biggest problem is for authentication (for now). After reading a LOT a documentation, It seems that the best solution is to use OpenID Connect to authenticate an user to retrieve a JWT that can by passed with the request to the microservices}. [\href{https://stackoverflow.com/questions/39134238/client-authentication-in-microservices-using-jwt-and-openid-connect}{{Taken from post 39134238}}]                                   &    \textbf{Security Discussion?} Yes  \newline \textbf{Security Words:} authentication; OpenID; JWT                             \\ \cline{2-3} 
                                        &        \say{If you would like to secure multiple services (deployed to the different app servers) with a common security provider, you should use Single-Sign-On approach}. [\href{https://stackoverflow.com/questions/58655233/how-to-spring-security-handle-all-modules-in-spring-boot}{{Taken from post 58655233}}]                            &         \textbf{Security Discussion?} Yes \newline \textbf{Security Words:} security; Single-Sign-On                        \\ \cline{2-3} 
                                       &     \say{You have done well identifying three bounded contexts, one for each domain and implemented in three microservices (MS). You are conforming to the best practices regarding DDD}.    [\href{https://stackoverflow.com/questions/44936115/authorization-in-microservice-architecture/44941512#44941512}{{Taken from post 44936115}}]                                &     \textbf{Security Discussion?} No                            \\ \hline 
\end{tabular}}
\end{table*}

\textbf{\textit {Main Study}}. In the main study, we annotated 12,393 paragraphs in \textit{goa}, \textit{eShopOnContainers}, \textit{microservices-demo}, \textit{deep-framework}, and \textit{light-4j} projects. Our pilot annotation experience taught us that the \textbf{SUM} item was not useful as it took a significant time of the annotators and had no positive effect on identifying security discussions. Therefore, we decided not to extract this item in the main study. Three annotators (three authors) annotated 12,393 paragraphs. 

Table \ref{sampleparagraph} shows examples of security (i.e., issue 647\footnote{\href{https://github.com/dotnet-architecture/eShopOnContainers/issues/647}{\url{https://bit.ly/3fPyBiC}\label{issue647}}} and issue 14\footnote{\href{https://github.com/networknt/light-4j/issues/14}{\url{https://bit.ly/3pk7Vt7}\label{issue14}}}) and non-security (i.e., issue 8\footnote{\href{https://github.com/networknt/light-4j/issues/8}{\url{https://bit.ly/3cg2KoW}\label{issue8}}}) discussions from GitHub, along with their annotations. The data collected from GitHub includes 1,602 security discussions and 10,791 non-security discussions.

\subsubsection{\textbf{Stack Overflow Data Collection}}\label{secSOdataset}

The number of security discussions collected from GitHub is relatively small compared to non-security discussions. This motivated us to explore Stack Overflow posts to identify more security discussions related to microservices systems. Moreover, the content, language usage, and sentence structure of Stack Overflow posts may differ {from} developer discussions captured by issue tracking systems in GitHub.

\textbf{\textit{Prepare Data}}. We extracted the posts with the ``microservice'' tag, {which resulted in} 5,655 posts. Being made up of numerous paragraphs, annotating such a large number of posts could take a great deal of time. Hence, we opted to choose a limited number of these posts, particularly those that increase the chance of identifying security discussions. To this end, we further filtered the 5,655 posts and only selected the posts that contained at least one of the following tags: \say{*secur*}, \say{*auth*}, \say{*safe*}, \say{*permiss*}, \say{*credential*}, \say{*access*}, \say{*identity*}, or \say{*jwt*}. 
This process resulted in 498 posts. Each post includes a question and a list of answers to the question \cite{zhang2019reading}. {Stack Overflow} provides mechanisms for users to append comments to both questions and answers. This enables them to have further discussions on the posted questions and answers. Hence, we included all 498 posts' questions, their questions' comments, answers, and answers' comments in our data analysis. In this way, we were exposed to valuable information regarding security discussions \cite{zhang2019reading}. The 498 posts contain 498 questions, 547 question’s comments, 470 answers, 520 answer's comments. We collected 4,884 paragraphs in the 498 posts using HTML tag $<$p$>$.

\textbf{\textit{Main Study}}. Three authors annotated the 4,884 paragraphs collected from Stack Overflow posts. One of them was also involved in annotating the data collected from GitHub (Section \ref{secgithubdataset}). 
These 4,884 paragraphs were annotated using the same approach discussed in Section \ref{secgithubdataset}. 
The data collected from Stack Overflow includes 3,211 security discussions (paragraphs) and 1,673 non-security discussions (paragraphs). Post 39134238 and post 58655233 in Table \ref{sampleparagraph} are two examples of the annotated security discussions from Stack Overflow. While post 39134238 is {a Stack Overflow question}, post 58655233 is a paragraph from the posted comments. Also, post 44936115 is a non-security discussion from the posted answers.

\subsubsection{\textbf{Dataset Reliability}}

Our dataset includes 17,277 labeled paragraphs. We randomly selected 581 paragraphs out of the 17,277 paragraphs (i.e., confidence level: 95\% and margin of error: 4\% \cite{kadam2010sample}). We asked the second author (i.e., cross-checker), who was not involved in the annotation process described in Sections \ref{secgithubdataset} and \ref{secSOdataset}, to annotate the 581 paragraphs. We calculated Cohen’s Kappa Coefficient to determine if there was an agreement between the 581 annotated paragraphs by the cross-checker and these 581 paragraphs annotated by other annotators. This resulted in a value of 0.82 for Cohen’s Kappa Coefficient, which is commonly perceived as \say{almost perfect agreement} \cite{mchugh2012interrater}.
The cross-checker held several meetings with the annotators to identify disagreements and solve them. The security-related words collected during the annotation process helped the cross-checker and the annotators to reach an agreement more straightforward. It is mainly because the security-related words enabled both sides to understand the reason behind selecting a given paragraph as a security discussion. {We make available the 581 paragraphs mentioned above of our dataset to enhance the reliability of our findings \cite{onlinedataset}.}

\section{Research Design}\label{secresearch}
The objective of this research is to help reduce the knowledge gap of practitioners in securing microservices systems. To this end, we formalized {three} research questions (RQs):

\begin{center}
\begin{tcolorbox}[colback=green!2!white,colframe=black!75!black]
\noindent \textit{\textbf{RQ1.} How do practitioners perceive security in microservices systems?}
\end{tcolorbox}
\end{center}

\noindent \textbf{Rationale}. The goal of \textbf{RQ1} is to collect software practitioners’ perceptions of security in microservices systems. {It also aims to realize if security discussions collected from past microservices systems can be useful for making security decisions during new microservices system development. \textbf{RQ1} is answered by a survey (Section \ref{secsurveyprocess}). We refer to it as ``preliminary survey''.}

\begin{center}
\begin{tcolorbox}[colback=green!2!white,colframe=black!75!black]
\noindent \textit{\textbf{RQ2.} Can we effectively identify security discussions automatically in developer discussions of microservices systems?}
\end{tcolorbox}
\end{center}

\noindent \textbf{Rationale}. Past and documented experiences of securing microservices systems are scarce (if any). Hence, security decisions in the context of microservices systems are often no well-informed. We believe that security knowledge (e.g., technical security advice) scattered in developer discussions can help bridge this gap \cite{lopez2019talking}. {As discussed in the Introduction section, manually identifying such security knowledge (e.g., security discussions) from developer discussions is not a straightforward task for practitioners.} \textbf{RQ2} aims to use and experiment with different ML and DL models (henceforth learning models) to distinguish security discussions from non-security discussions automatically (Section \ref{secexp}). 
\begin{center}
\begin{tcolorbox}[colback=green!2!white,colframe=black!75!black]
\noindent {\textit{\textbf{RQ3.} Do practitioners find the automated identification of security discussions in microservices systems useful in practice? If so, how?}}
\end{tcolorbox}
\end{center}

\noindent {\textbf{Rationale}. The automated identification of security discussions would be helpful in practice only if the results produced by the ML/DL models (\textbf{RQ2}) are perceived useful by microservices practitioners. \textbf{RQ3} focuses on the results produced by the best-performing learning model and employs a survey (we call it ``validation survey'') to assess whether practitioners perceive those results useful in practice. If so, the validation survey further investigates how practitioners can use the produced results in practice (Section \ref{secvalidationresulst}).}

\subsection{\textbf{Preliminary Survey (RQ1)}} \label{secsurveyprocess}

We used an online survey to solicit practitioners' perspectives on security in microservices systems (\textbf{RQ1}). Here we demonstrate the process of conducting the survey.

\textbf{Protocol}. Following the guidelines suggested by \cite{shull2007guide}, we designed a short survey\footnote{\url{http://tiny.cc/cj0vsz}} to collect software practitioners' perspectives on security in microservices systems. Our survey was anonymous to motivate more practitioners to participate and encourage them to provide honest answers \cite{johnson2019effect}. The survey was hosted on Google Forms. In total, the survey had nine questions, including three demographic questions (e.g., \say{how many years have you {been} involved in microservices system development?}), five Likert scale questions, and one open-ended question. We made all questions except the open-ended question mandatory. At the preamble of the survey, we defined \say{security discussion} to avoid misinterpretations among the participants. The participants were asked to show how strongly they agree or disagree with five statements regarding security in microservices systems. The statements were designed based on the literature \cite{bogner2019microservices}, \cite{ghofrani2018challenges}, \cite{sun2015security}, \cite{yarygina2018overcoming} and rated on a five-point Likert scale (\say{strongly agree = 5} to \say{strongly disagree = 1}). The statements are:
\begin{itemize}
\item Statement 1. \say{\textit{The microservices architecture (MSA) style brings unique security challenges}}.
\item Statement 2: \say{\textit{It is more challenging to address security in microservices systems compared to traditional service-oriented systems or monolithic systems}}.
\item Statement 3. \say{\textit{There is a knowledge gap among software practitioners on how to secure microservices systems}}.
\item Statement 4. \say{\textit{Software practitioners will make better security decisions while developing microservices systems if they are acquainted with the security discussions collected from the past microservices systems}}.
\item Statement 5. \say{\textit{A tool that can automatically collect security discussions would be useful}}.
\end{itemize}
The open-ended question allowed the participants to share any general comments about security concerns in microservices systems.

\textbf{Participants}. In Section \ref{secconsult}, we found 167 open-source projects that had the potential to be designed based on the MSA style. Although we finally found that not all of these 167 projects followed the MSA style, many of these projects developed frameworks, tools, or libraries to support the development of microservices systems. We collected the publicly available email addresses of 868 software practitioners who contributed to these 167 open-source projects. We emailed all of the 868 software practitioners and invited them to fill out the survey. We got 67 responses (response rate: 7.7\%).

\textbf{Data Analysis}. We employed descriptive statistics to analyze the demographic and Likert scale questions. We used open coding and constant comparison methods in grounded theory to analyze the qualitative responses to the open-ended question \cite{glaser1968discovery}.

\subsection{\textbf{Experiments (RQ2)}}\label{secexp}
{To automatically identify security discussions, we developed a wide range of ML and DL based models. We applied these models to the dataset described in Section \ref{securitydiscussions}.}

\subsubsection{\textbf{Pre-processing}}\label{secPreprocess}

We applied four pre-processing steps to remove possible noises from our dataset and polish the dataset.

\textit{Step 1. Removing useless characters and irrelevant words}. Developers usually use informal language when they share their opinions in issue tracking systems or answer Stack Overflow questions \cite{tao2020identifying}. For example, they may utilize emoji icons to indicate their feeling about a decision or a discussion. Links may also be added to developer discussions to provide, for example, background information \cite{hata20199}. Emoji icons, links, emails, and useless characters (e.g., `/', `*') in developer discussions do not provide credible information for Natural Language Processing (NLP) techniques. Even they may decrease the performance of classifiers. We used regular expressions in Python to remove all of them from our dataset.

{\textit{Step 2. Converting short form of words to their full forms}}. Some developers prefer to use abbreviations (e.g., \say{wouldn't}, \say{we'll}, and \say{can't}) when communicating their thoughts \cite{li2020automatic}. We identified such words and converted them to their full forms.

{\textit{Step 3. Removing stop words}}. 
Removing stop words could positively affect the performance of learning classifiers, such as reducing noises and increasing recall value \cite{silva2003importance}. Therefore, we removed stop words from our dataset using Neural Language Toolkit (NLTK) \cite{bird2009natural}.

{\textit{Step 4. Stemming process}}. 
Stemming is a widely used normalization technique in NLP. Stemming removes the derivational affixes of a word and converts the word to its common base form (e.g., reducing \say{going} to \say{go}) \cite{porter1980algorithm}, \cite{manning2008introduction}.

Research has shown that simple stemming techniques (e.g., the Porter stemming technique) can improve recall without significantly degrading precision \cite{kraaij1996viewing}. We used the Porter stemming technique to convert words in the dataset to their base forms.

\subsubsection{\textbf{Machine Learning Models}}

We used several ML algorithms, which are widely used in software engineering research \cite{viviani2019locating}, \cite{aniche2020effectiveness}, \cite{bao2019large}, \cite{yu2020identifying}, to predict security discussions. More specifically, we used (1) Random Forest (RF) \cite{breiman2001random}, (2) Decision Tree (DT) \cite{quinlan1986induction}, (3) Support Vector Machine based on Linear Kernel (SVM-LR) \cite{hsu2003practical}, \cite{cortes1995support}, and (4) Extreme Gradient Boosting (XGBoost) \cite{chen2016xgboost}. We used three text feature extraction techniques to extract features from paragraphs in our dataset, which can be used as input for ML algorithms. We briefly introduce them:

\begin{itemize}
\item \textbf{BoW}. BoW (Bag of Word) aims to count the occurrence frequency of a unique word in a document \cite{mccallum1998comparison}, \cite{joachims1998text}. This technique does not consider other language aspects of the sentence, such as grammar and word order.

\item \textbf{TF-IDF}. In contrast to BoW, TF-IDF (Term Frequency-Inverse Document Frequency) is a more complicated feature engineering method that estimates each word's importance in a corpus of documents \cite{ramos2003using}. A word's weight is obtained by considering two factors: the number of appearing the word is in the document (TF) and the inverse document frequency (IDF) of the word across the corpus.
\item \textbf{GloVe}. Word embedding techniques represent each word in a corpus (e.g., a set of text documents) with a word vector \cite{turian2010word}, \cite{ollagnier2019classification}. Research has shown that word embedding techniques can enhance the performance of ML classifiers \cite{ghannay2016word}. GloVe (global vectors for word representation), as one of the most-used embedding techniques, maps the words to a meaningful space and creates a semantic similarity between them \cite{pennington2014glove}.
GloVe uses a pre-trained word vector embedding composed of 300-dimensional vectors with 6B tokens and 400K vocabularies \cite{glovStandoford}.
\end{itemize}

\begin{figure*}[h]
	\centering
		\includegraphics[scale=.61]{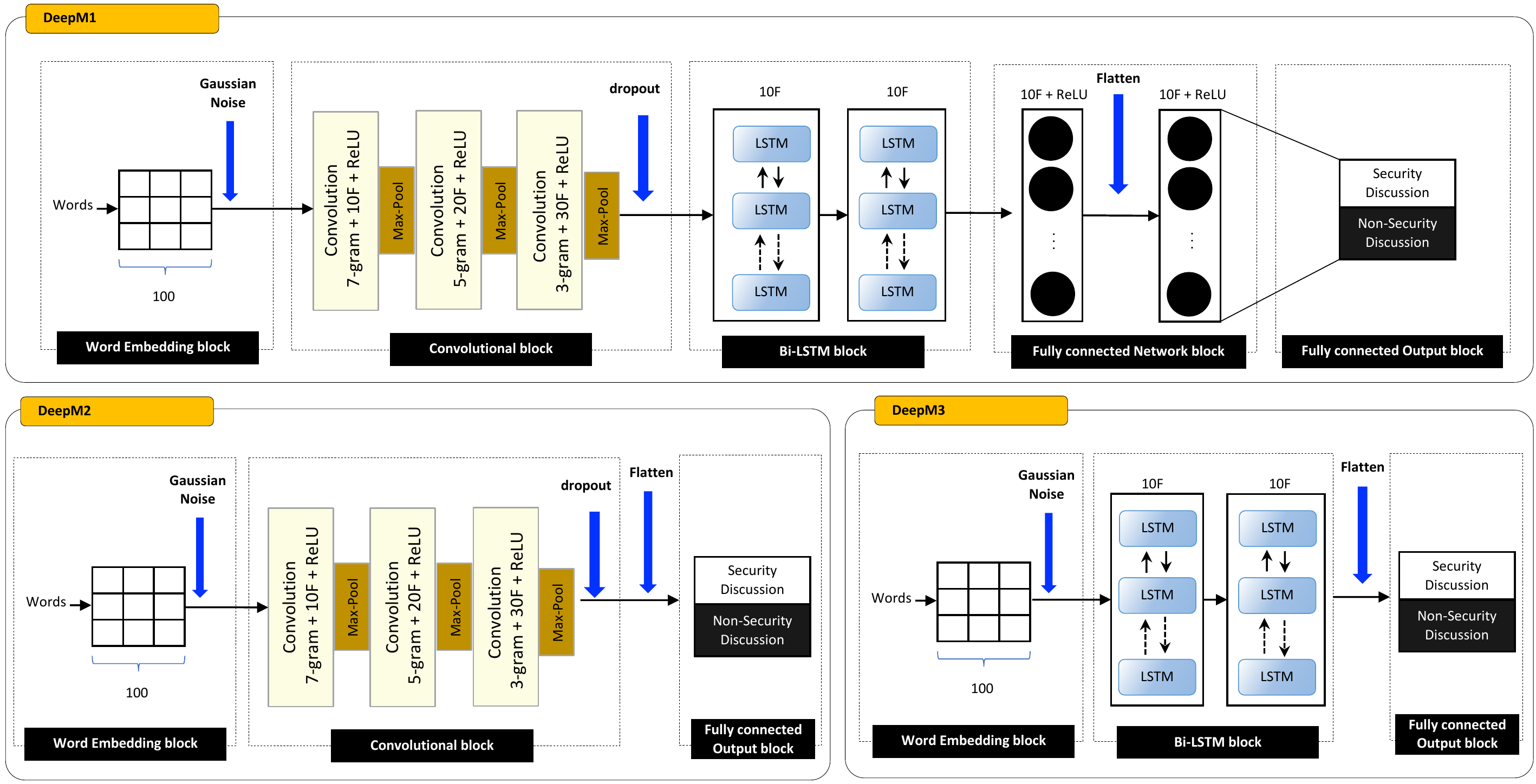}
	\caption{An overview of DeepM1, DeepM2, and DeepM3}
	\label{FIG:deepmodels}
\end{figure*}

\subsubsection{\textbf{Deep Learning Models}}\label{secDL}

We developed three DL models (DeepM1, DeepM2, and DeepM3) to automatically classify security discussions from developer discussions of microservices systems. Figure \ref{FIG:deepmodels} shows an overview of DeepM1, DeepM2, and DeepM3. Below, we describe these DL models.

\textbf{\textit{DeepM1.}}\label{secCBLSTM} DeepM1 includes five blocks, in which three CNN (Convolutional Neural Networks) layers and two Bi-LSTM (Bi-directional Long Short-Term Memory) networks play significant roles (i.e., key building blocks) \cite{kalchbrenner2014convolutional}, \cite{zhao2017learning}, \cite{schuster1997bidirectional}, \cite{lecun2015deep}. While a wide range of deep network types can be used to develop a DL model, we opted to use CNN and Bi-LSTM in DeepM1 as they tend to be very popular in the text processing tasks \cite{zhang2018deep}, \cite{basiri2020abcdm}. The popularity of CNN stems from its ability to learn local patterns in the textual data. LSTM can better deal with long-term dependencies in the input data than its competitors, such as recurrent neural networks (RNNs) \cite{song2019attention}.

\ding{202} \textbf{Input and Word Embedding block.} In the first place, we assign a unique integer to each unique word in our dataset. The unique numbers are assigned to the unique words {so} that a smaller number shows that the word is more frequent in our dataset. Then, a corpus of the unique words and their assigned unique numbers is created.

Then, for each paragraph in our dataset, its words will be replaced by the unique numbers produced in the previous step \cite{basiri2020bidirectional}. This process converts each paragraph to a sequence of numbers. The size of each sequence {is equal to} its corresponding paragraph length. However, the lengths of all produced sequences need to be equal. We use the zero-padding technique \cite{gulli2017deep}, \cite{mastery} to equalize the lengths of all the sequences (i.e., the length of all sequences is set to 100). In the first layer of DeepM1, we choose a word embedding layer \cite{gulli2017deep}, \cite{mastery} and set the embedding dimension to 100. The embedding layer maps the word indexes of the corpus to their dense vector representations \cite{basiri2020bidirectional}, \cite{lopez2018hybridizing}. To reduce overfitting, a GaussianNoise is applied immediately after the word embedding \cite{gulli2017deep}, \cite{mastery}.

\ding{203} \textbf{Convolutional block.} This block performs feature extraction and includes three convolution layers and three pooling layers. The first layer of this block takes the embedding vectors as input and converts them to a feature map. The feature map captures all the relevant information about the embedding vectors.
DeepM1 uses three convolutional networks with the kernel size and feature map of 7-gram and 10, 5-gram and 20, and 3-gram and 30.
Hence, to increase the performance of DeepM1, each of these convolutional networks is configured based on the Same-Padding \cite{wiranata2018investigation} and ReLU (Rectified Linear Unit) \cite{ide2017improvement} functions. 
This block also employs max-pooling to merge the output of the convolutional networks \cite{basiri2020novel}. To this end, {a max-pooling operation} is applied with a factor of 2 after each convolutional network to identify and select the most important features \cite{collobert2011natural}.
In the last step, a 0.25 dropout is added to handle the overfitting \cite{srivastava2014dropout}.

\ding{204} \textbf{Bi-LSTM block.} This block takes the feature map produced in the previous block and transforms it into a compressed representation \cite{ml}. Recent studies \cite{zhou2016text}, \cite{basiri2020abcdm}, \cite{basiri2020novel} have shown that the DL models based on Bi-LSTM and CNN could obtain promising performance in the text classification tasks. Bi-LSTM is considered as an appropriate option when the sequence is long enough \cite{rao2018lstm}. 
Bi-LSTM cells contain two LSTM layers in opposite directions (i.e., forward and backward), which can adequately deal with the sequential modeling problem \cite{basiri2020abcdm}.
Hence, we utilize two layers of Bi-LSTM with {a dimensionality factor of 10.}

\ding{205} \textbf{Fully Connected Network block.} We also use two dense layers (i.e., fully-connected networks), in which each of them has 10 neurons. A flatten is placed between these two dense layers to merge 10 neurons into another column matrix \cite{gulli2017deep}. The ReLU function is embedded in fully connected networks.

\ding{206} \textbf {Fully Connected Output block.} The previous block's output is received by a fully connected output layer, which classifies a paragraph into a security discussion or non-security discussion.

\textbf{\textit{DeepM2.}}\label{secCNN}
This model uses three blocks of DeepM1: input and word embedding block, convolutional block, and fully connected output block. It should be noted that a flatten is applied between the last CNN layer and the fully connected output layer.

\textbf{\textit{DeepM3.}}\label{secBLSTM}
This model utilizes blocks 1, 3, and 5 of DeepM1 (i.e., input and word embedding block, Bi-LSTM block, and fully connected output block). After the second Bi-LSTM layer, a flatten is applied.

{The rationale for choosing convolution layers in DeepM1 and DeepM2 is to extract different textual features. Specifically, using three convolution layers in DeepM1 and DeepM2, we can extract n-grams features. LSTM layers, on the other hand, are used in DeepM1 and DeepM3 to consider context information in the text. To investigate the effect of using these two types of layers, we examined them independently in DeepM2 and DeepM3 and jointly in DeepM1.}
Note that we optimize all deep learning models using Adam algorithm \cite{kingma2014adam} with a learning rate of 1e-3 and a decay rate of 1e-6. Each DL model's batch size is set to 64, and its network is trained for 100 epochs with early stopping.

\subsubsection{\textbf{Performance Evaluation}} \label{secEvaluation}
To appraise  {the ML/DL models}, we need a confusion matrix containing four factors: False Positive (FP), True Positive (TP), True Negative (TN), False Negative (FN). FP indicates the number of non-security discussions classified as {security discussions} TP signifies the number of security discussions that are accurately classified, and TN represents the number of non-security discussions that are precisely classified.
FN represents the number of classes of actual security discussion paragraphs that are classified as non-security discussion paragraphs. 
Precision, recall, and F1-score are common metrics to compare the outputs of learning models (e.g., \cite{bettaieb2019decision}, \cite{aniche2020effectiveness}, \cite{yu2020identifying}, \cite{abualhaija2019machine}). Precision holds the percentage of identified security discussions that are actually security discussions \cite{abdalkareem2020machine} (i.e., $Precision= \frac{TP}{TP+FP}$). Recall displays the ratio between the number of accurately detected security discussions and the total number of security discussions (i.e., $
Recall= \frac{TP}{TP+FN}$). F1-score is computed by the combination of the recall metric and precision metric \cite{pedregosa2011scikit} (i.e., $F1-score = \frac{2*Precision*Recall}{Precision + Recall}$). 

Other studies have recommended or used AUC \cite{bradley1997use}, \cite{fawcett2006introduction} as an evaluation metric to measure the performance of classifiers \cite{viviani2019locating}, \cite{bao2019large}, \cite{lessmann2008benchmarking}, \cite{tantithamthavorn2018experience}. In our context, the AUC metric calculates the likelihood that a prediction model will rank a randomly selected security discussion (i.e., TP) higher than a randomly chosen non-security discussion (i.e., FP) \cite{viviani2019locating}.
G-mean is another metric that can be used to measure the quality of binary classifications \cite{le2020puminer}, \cite{peters2017text}. Specifically, G-mean metric is highlighted as the best metric for imbalanced classes that consist of both positives and negatives, especially if classification errors being considered \cite{luque2019impact} (i.e., $G-mean = \frac{\sqrt{TP*TN}}{\sqrt{(TP+FN)*(TN+FP)}}$). All metrics range from 0 to 1. It is obvious that 1 represents the best performance, and 0 value shows the worst performance prediction model.

We utilize 10-fold cross-validation technique for the performance evaluation of ML/DL models. The whole dataset (i.e., {17,277 paragraphs collected from both GitHub and Stack Overflow} (Section \ref{secBuilddataset})) is divided randomly into ten separate folds. The divided data is evaluated 10 times. Each time, 10\% of data (i.e., one fold) is selected for performance evaluation, and the rest (i.e., 90\%) is used to train the system. In this study, ML/DL models are set to a fixed seed for reproducibility \cite{mastery}, \cite{zimmerman2018improving}.

\subsubsection{Experiment Setup and Implementation}\label{secExpImp}
All collected data were stored in a MySQL database by using a Python script. We implemented all experiments using scikit-learn\footnote{\href{https://scikit-learn.org/stable/}{\url{https://bit.ly/2S7vtFB}}} and Keras\footnote{\href{https://keras.io}{\url{https://bit.ly/3cf2SVU}}} libraries in Python. We executed the experiments on Google Colab with an Intel Xeon CPU and 13GB of RAM.

\subsection{Validation Survey (RQ3)}
\label{validationsurvey}
{We conducted an online survey\footnote{\url{https://bit.ly/3u4oU4G}} to seek and analyze the perceptions of practitioners about the usefulness of the results produced by the learning models (RQ2). We detail the execution of the survey in the following.}

{\textbf{Protocol.} We designed an anonymous survey hosted on the Qualtrics platform with 11 questions. In the servery preamble, we described the problem statement, our proposed approach, and the survey's objective. In the problem statement, we explained why security decisions in microservices systems are often not well informed. Then, we described that we developed an approach that can automatically identify security discussions with high accuracy from a large number of issue discussions in microservices systems to (partially) address this problem. We also defined “security discussion” to avoid misinterpretations. The participants were shown the distinguished security discussions from non-security discussions using our best learning model (i.e., DeepM1) in issue 303 from the Moleculer project and in issue 803 from the Goa project.} 

{Out of the 11 survey questions, two questions were optional, and the rest was compulsory. The survey included three demographic questions, in which one of them was optional (i.e., which country do you currently work in?), seven Likert scale questions, and one open-ended question. The optional open-ended question was used to seek the participants' suggestions to improve our approach (i.e., what improvements to our approach would further help develop secure microservices systems?). We asked the participants to rate the level of their agreement or disagreement (``strongly agree = 5'' to ``strongly disagree = 1'') with the following seven statements. We consulted the (microservices) security-related literature} (e.g., \cite{palacio2019learning}, \cite{yarygina2018overcoming}, \cite{lopez2019anatomy}, \cite{esposito2016challenges}) to design these seven statements.
\begin{itemize}

\item {Statement 1. ``\textit{The approach is useful because security discussions identified by the approach convey meaningful and important security information}''.}

\item {Statement 2. ``\textit{The approach is useful because security discussions identified by the approach can be used to make informed security decisions in the future or refine the existing sub-optimum security decisions}''.}

\item {Statement 3. ``\textit{The approach is useful because I, as a practitioner, can find useful materials in a reasonable time slot from security discussions identified by the approach}''.}

\item {Statement 4. ``\textit{The approach is useful because security discussions identified by the approach can help us identify security-critical issues (e.g., security bugs, security mistakes, etc.) faster in our systems than if we do it manually}''.}

\item {Statement 5. ``\textit{The approach is useful because security discussions identified by the approach may contain incoming security-sensitive bug reports, with which they can be readily identified and more effectively prioritized and resolved}''.}

\item {Statement 6. ``\textit{The approach is useful because security discussions identified by the approach can provide cues/hints to trace backward and forward to security-critical artifacts (e.g., codes, services) and features}''.}

\item {Statement 7. ``\textit{The approach is useful because the security discussions identified by the approach could be beneficial for those who have little security experience in microservices or recently joined a microservices project as they can quickly access and learn security solutions and avoid common security mistakes}''.}
\end{itemize}

{\textbf{Participants.} We recruited the participants in three ways. First, we invited all the 868 practitioners who were also asked to fill out the survey designed for RQ1. Then, we advertised our survey via social networks, such as LinkedIn (e.g., microservices groups on LinkedIn). Finally, we approached the relevant microservices practitioners on LinkedIn. We thoroughly analyzed their profiles and invited them directly via email or by sending a LinkedIn message. We received 51 responses finally. Due to the complexity of the recruitment process, we were not able to calculate the response rate.}

{\textbf{Data Analysis.} We used descriptive statistics to analyze the responses to the demographic and Likert scale questions. The responses to the open-ended question were analyzed using open coding and constant comparison \cite{glaser1968discovery}.}

\section{\textbf{Findings}}\label{secfindings}

This section first reports the results of the preliminary survey to answer \textbf{RQ1} (Section \ref{secsurvfinding}). Then, we present the findings of the experiments conducted to answer \textbf{RQ2} (Section \ref{secexpfinding}). Finally, we report the findings of the validation survey to respond to \textbf{RQ3} (Section \ref{secvalidationresulst}). 

\begin{figure*}[t]
	\centering
		\includegraphics[scale=.75]{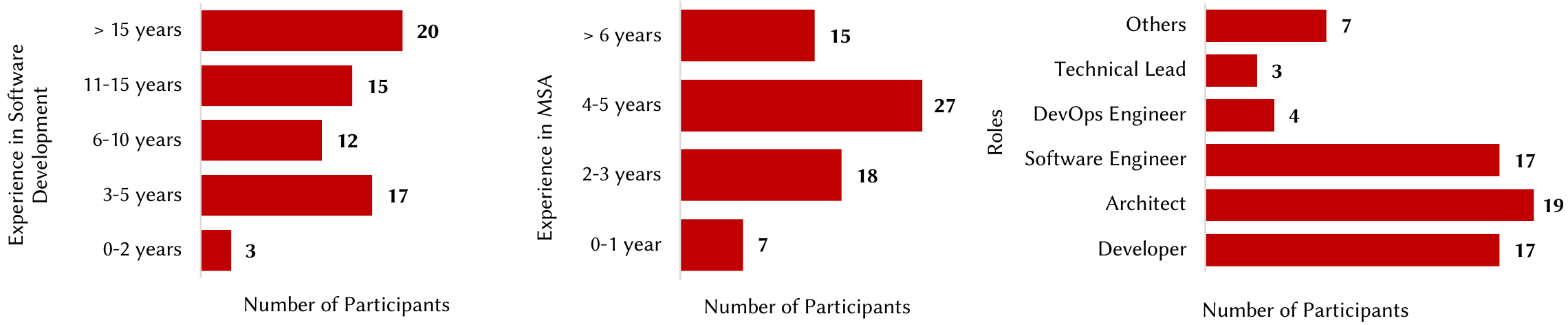}
	\caption{The preliminary survey: participants demographic data (n=67)}
	\label{FIG:demoResult}
\end{figure*}

\begin{figure*}[t]
	\centering
		\includegraphics[scale=.75]{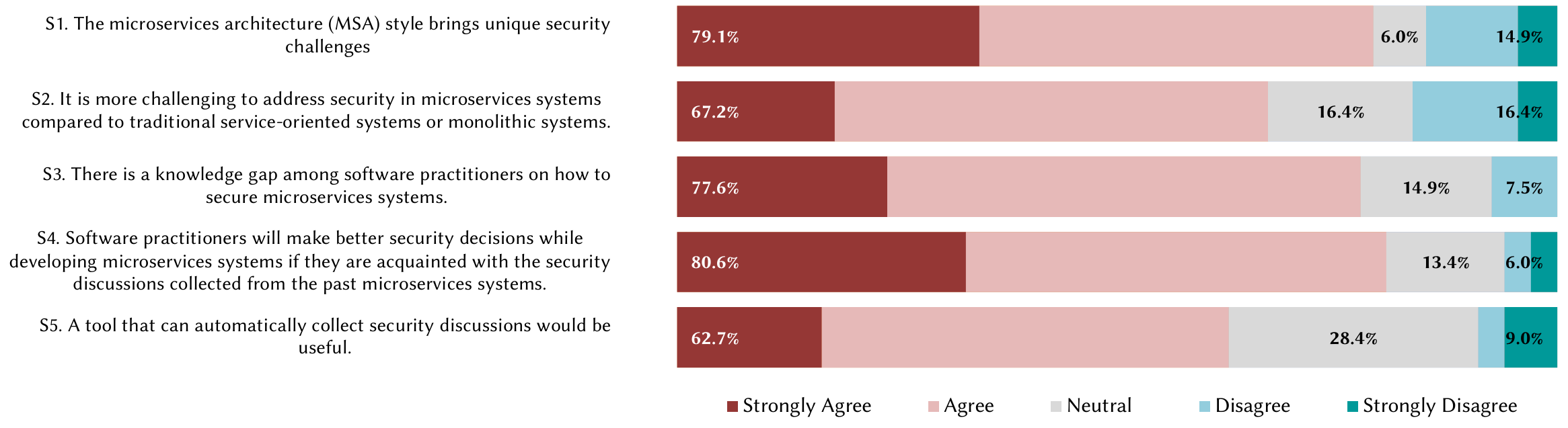}
	\caption{The preliminary survey: responses to the statements regarding security in microservices systems (n=67)}
	\label{FIG:likerResult1}
\end{figure*}

\subsection{\textbf{Practitioners' Perspective on Security in Microservices Systems (RQ1)}}\label{secsurvfinding}

Figure \ref{FIG:demoResult} represents an overview of the participants’ demographics. More than half of the participants (52.3\%, 35 out of 67) had developed software for at least 11 years. Of 67 participants, only 3 had less than two years of software development experience. The majority of the participants (62.7\%) were involved in microservices system development for more than 4 years. The dominant roles who participated in our survey were architects (19), followed by developers (17), software engineers (17), DevOps engineers (4), and technical leads (3). 

Through 5 {Likert} scale questions, we sought the opinions of the respondents about security in microservices systems. As shown in Figure \ref{FIG:likerResult1}, the majority of them (79.1\%) (strongly) agreed that the MSA style brings a unique challenge (statement S1). A participant pointed out the challenging nature of establishing security in microservices systems as follows: \say{\textit{When we architect microservices, security is among the top things we nail down, i.e., will the microservice get exposed to the public, will the microservice perform authentication and authorization or does it trust all requests coming in}} (Developer).

On the other hand, only 10 (14.9\%) participants (strongly) disagreed with this statement, and {the rest (4 respondents, 6\%)} took a neutral position. When we asked \say{\textit{it is more challenging to address security in microservices systems compared to traditional service-oriented systems or monolithic systems}}, the participants rated this statement (S2) \textit{strongly agree} (17.9\%), \textit{agree} (49.3\%), \textit{neutral} (16.4\%), \textit{disagree} (11.9\%), and \textit{strongly disagree} (4.5\%). Our survey results reveal the knowledge gap amongst practitioners in securing microservices systems as 52 out of 67 respondents (77.6\%) answered statement S3 as \textit{strongly agree} or \textit{agree}. No one rated statement S3 as \textit{strongly disagree}. An architect believed that \say{\textit{helping small teams develop secure microservices is important}}. Another survey respondent shared his concern about the knowledge gap among practitioners in securing microservices deployed in a container: \say{\textit{most people who implement microservices will deploy their services inside a container, and I heard/read that some of them assume that using container will by default is secure, this is misleading}} (Software Engineer). A software engineer elaborated on this and said, \say{\textit{I think it would be a good idea to collect how people secure their microservices inside a container, did they use the public image? Did they inspect/scan image before using it, and so on}}.

The respondents were asked to indicate to what extent they agree or disagree that the security discussions collected from the previous {microservices systems can help practitioners make better security decisions} while developing such systems (statement S4). The majority of the participants (80.6\%) strongly agreed or agreed with this statement (S4). Less than 10\% of the respondents rated statement S4 as \textit{strongly disagree} or \textit{disagree}. Our participants also confirmed the usefulness of the potential automated tools that can collect and detect security discussions from developer discussions (i.e., 16.4\% strongly agree and 46.3\% agreed with statement S5). While only 9\% of the respondents (strongly) disagreed that such tools can be useful, statement S5 received the highest neutral responses (19 out of 67, 28.4\%). One participant pointed out: \say{\textit{Not sure how a tool can help in making a system secure. I believe Guidelines on Security can be a better alternative}} (Developer). 

\renewcommand{\arraystretch}{1.2}
\begin{table*}[h]
\centering
\caption{Precision, Recall, F1-Score, AUC and G-mean for ML and DL models. The best results of each metric are grayed.}\label{mldlresults}
{\small
\begin{tabular}{clccccc}
\hline 
         \multicolumn{1}{c}{}                          & \multicolumn{1}{l}{\textbf{Models}}  & \multicolumn{1}{c}{\textbf{Precision}} & \multicolumn{1}{c}{\textbf{Recall}} & \multicolumn{1}{c}{\textbf{F1-score}} & \multicolumn{1}{c}{\textbf{AUC}}   & \multicolumn{1}{c}{\textbf{G-mean}} \\ \hline

\multirow{12}{*}{Machine learning} & BoW + RF         & 92.03              & 77.77           & 84.30             & 87.58          & 87.03           \\  
                                   & BoW + DT         & 83.12              & 80.20           & 81.63             & 86.95          & 86.69           \\ 
                                   &  BoW + XGBoost    &    \cellcolor{whitesmoke} 93.43      & 71.20           & 80.82             & 84.63          & 83.56           \\ 
                                   & BoW + SVM-LR     & 86.92              & 83.08           & 84.96             & 89.12          & 88.92           \\  
                                   & TF-IDF + RF      & 91.93              & 78.54           & 84.71             & 87.94          & 87.43           \\ 
                                   & TF-IDF + DT      & 82.91              & 81.45           & 82.17             & 87.48          & 87.27           \\ 
                                   & TF-IDF + XGBoost & 93.29              & 71.30           & 80.83             & 84.66          & 83.60           \\ 
                                   & TF-IDF + SVM-LR  & 91.65              & 78.47           & 84.55             & 87.85          & 87.35           \\  
                                   & GloVe + RF       & 89.18              & { 49.64}     & 63.77             & 73.65          & 69.63           \\ 
                                   & GloVe + DT      & { 52.90}        & 55.72           & { 54.28}       & { 68.27}    & { 67.11}     \\ 
                                   & GloVe + XGBoost  & 82.46              & 59.48           & 69.11             & 77.30          & 75.22           \\ 
                                   & GloVe + SVM-LR   & 82.19              & 59.44           & 68.99             & 77.23          & 75.16           \\ \hline
  
\multirow{3}{*}{Deep Learning} & DeepM1      & 86.73              &   \cellcolor{whitesmoke} 84.25     &  \cellcolor{whitesmoke}85.47   &  \cellcolor{whitesmoke}89.63  & \cellcolor{whitesmoke}  89.07 \\ 
                                   & DeepM2      & 80.16              & 79.11           & 79.64             & 85.77          & 85.51           \\ 
                                   & DeepM3      & 84.01              & 82.36           & 83.18             & 88.15          & 87.96           \\ \hline 
                               \multicolumn{2}{c}{Average}          & 84.86              & 72.80           & 77.89             & 83.75          & 82.77   \\ \hline        
\end{tabular}}
\end{table*}

\begin{center}
\begin{tcolorbox}[colback=gray!5!white,colframe=black!75!black]
\textbf{RQ1 Summary:} \textit{{Although the preliminary survey participants perceive securing microservices systems as a unique challenge, they find collecting and leveraging developer security discussions scattered in existing microservices systems useful for making security decisions. Furthermore, a tool that automatically collects such security discussions is deemed useful by the participants.}}
\end{tcolorbox}
\end{center}

\subsection{\textbf{Learning to Identify Security Discussions (RQ2)}}\label{secexpfinding}

To answer RQ2, we evaluated 15 ML/DL models (See Section \ref{secexp}). ML models include Random Forest, Decision Tree, XGBoost, and SVM-LR algorithms based on three text feature selection techniques: BoW, TF-IDF, and GloVe. Three DL models (DeepM1, DeepM2, and DeepM3) use the word embedding technique. 
It should be noted that the training process of DL models is supposed to stop if there are no changes observed in the best accuracy in the last 10 epochs (i.e., early stopping).
Table \ref{mldlresults} shows the performance evaluation of ML and DL models using 10-fold cross-validation technique over 17,277 paragraphs.
In Table \ref{mldlresults}, the grayed values show the best results for each metric. As shown in Table \ref{mldlresults}, {ML and DL models} achieve, on average, a precision of 84.86\%, recall of 72.80\%, F-score of 77.89\%, AUC of 83.75\%, and G-mean of 82.77\%. Among all ML and DL models, DeepM1, as a DL model, performs the best in all metrics except in the precision metric. XGBoost with BoW achieved the best performance in the precision metric (precision: 93.43\%). DeepM1, DeepM3, Decision Tree with TF-IDF and BoW, and SVM-LR with BoW achieved a performance of more than 80\% in all metrics. The worst results were obtained when GloVe was used as the feature selection technique.

\begin{table*}
\centering
\caption{Performance of fifteen learning models on two unseen datasets. \textit{GitHubGen dataset}: 500 paragraphs from 2 microservices systems; \textit{StackExGen dataset}: 77 paragraphs from 7 posts of Security Stack Exchange.}\label{generalization}
{
\resizebox{12.5cm}{!}{\begin{tabular}{clccccc}
\hline 
\multicolumn{1}{c}{\textbf{Unseen Datasets}}                                                                & \multicolumn{1}{l}{\textbf{Models}} & \multicolumn{1}{c}{\textbf{Precision}} & \multicolumn{1}{c}{\textbf{Recall}} & \multicolumn{1}{c}{\textbf{F1-score}} & \multicolumn{1}{c}{\textbf{AUC}} & \multicolumn{1}{c}{\textbf{G-mean}} \\ \hline
\multirow{15}{*}{GitHubGen}                                                             & BoW+RF          & 96.34              & 55.63           & 70.54             & 77.39        & 74.27           \\ 
                                                                                          & BoW+DT          & 89.52              & 66.19           & 76.11             & 81.56        & 80.10           \\ 
                                                                                          & BoW+XGBoost     & \cellcolor{whitesmoke} 100                & 46.47           & 63.46             & 73.23        & 68.17           \\ 
                                                                                          & BoW+SVM-LR      & 80.39              & 57.74           & 67.21             & 76.07        & 73.83           \\ 
                                                                                          & TF-IDF+RF       & 96.20              & 53.52           & 68.78             & 76.34        & 72.85           \\ 
                                                                                          & TF-IDF+DT       & 85.59              & 71.12           & 77.69             & 83.18        & 82.30           \\ 
                                                                                          & TF-IDF+XGBoost  & 98.44              & 44.36           & 61.17             & 72.04        & 66.51           \\ 
                                                                                          & TF-IDF+SVM-LR   & 95.29              & 57.04           & 71.37             & 77.96        & 75.10           \\ 
                                                                                          & GloVe+RF        & 84.62              & 15.49           & 26.19             & 57.18        & 39.14           \\ 
                                                                                          & GloVe+DT        & 36.59              & 31.69           & 33.96             & 54.95        & 49.78           \\ 
                                                                                          & GloVe+XGBoost   & 65.62              & 29.57           & 40.78             & 61.71        & 52.68           \\ 
                                                                                          & GloVe+SVM-LR    & 61.64              & 31.69           & 41.86             & 61.93        & 54.04           \\ 
                                                                                          & DeepM1          & 86.71              & {\cellcolor{whitesmoke} 78.16 }          & \cellcolor{whitesmoke} 82.22             & \cellcolor{whitesmoke} 86.71        & \cellcolor{whitesmoke} 86.28           \\ 
                                                                                          & DeepM2          & 82.30              & 75.35           & 78.67             & 84.46        & 83.97           \\
                                                                                          & DeepM3          & 85.88              & 51.40           & 64.31             & 74.02        & 70.48           \\  \hline 
                                                                                                   \multicolumn{2}{c}{Average}          &     78.90          & 51.03           &   61.62           &     73.25    &    68.63        \\ \hline 
\multirow{15}{*}{StackExGen}                                                            & BoW+RF          & 96.08              & 77.77           & 85.96             & 81.74        & 81.64           \\ 
                                                                                          & BoW+DT          & 98.08              & 80.95           & 88.70             & 86.90        & 86.70           \\ 
                                                                                          & BoW+XGBoost     & 97.87              & 73.01           & 83.64             & 82.93        & 82.34           \\ 
                                                                                          & BoW+SVM-LR      & 98.11              & 82.53           & 89.66             & 87.69        & 87.54           \\ 
                                                                                          & TF-IDF+RF       & 98.00              & 77.77           & 86.73             & 85.31        & 84.98           \\ 
                                                                                          & TF-IDF+DT       & 96.15              & 79.36           & 86.96             & 82.53        & 82.47           \\ 
                                                                                          & TF-IDF+XGBoost  & 97.87              & 73.01           & 83.64             & 82.93        & 82.34           \\ 
                                                                                          & TF-IDF+SVM-LR   & 98.08              & 80.95           & 88.70             & 86.90        & 86.70           \\ 
                                                                                          & GloVe+RF        & 93.94              & 49.20           & 64.58             & 67.46        & 64.94           \\ 
                                                                                          & GloVe+DT        & 93.62              & 69.84           & 80.00             & 74.20        & 74.07           \\ 
                                                                                          & GloVe+XGBoost   & 95.00              & 60.31           & 73.79             & 73.01        & 71.90           \\ 
                                                                                          & GloVe+SVM-LR    & 90.48              & 60.31           & 72.38             & 65.87        & 65.63           \\ 
                                                                                          & DeepM1          & \cellcolor{whitesmoke} 98.18              & 85.71           & 91.51             & \cellcolor{whitesmoke} 89.28        & \cellcolor{whitesmoke} 89.21           \\ 
                                                                                          & DeepM2          & 96.55              & \cellcolor{whitesmoke} 88.88           & \cellcolor{whitesmoke} 92.56             & 87.30        & 87.28           \\ 
                                                                                          & DeepM3          & 98.14              & 84.12           & 90.59             & 88.49        & 88.38           \\  \hline
                                                                                                  \multicolumn{2}{c}{Average}          &   96.41           &     74.91       &        83.96      &    81.50     &      81.07      \\ \hline  
\multirow{15}{*}{\begin{tabular}[c]{@{}c@{}}GitHubGen\\ + \\ StackExGen\end{tabular}} & BoW+RF          & 97.58              & 59.02           & 73.56             & 79.10        & 76.51           \\ 
                                                                                          & BoW+DT          & 91.03              & 69.26           & 78.67             & 82.75        & 81.64           \\ 
                                                                                          & BoW+XGBoost     & \cellcolor{whitesmoke} 99.12              & 54.63           & 70.44             & 77.18        & 73.81           \\ 
                                                                                          & BoW+SVM-LR      & 86.45              & 65.36           & 74.44             & 79.86        & 78.53           \\ 
                                                                                          & TF-IDF+RF       & 95.56              & 62.92           & 75.88             & 80.65        & 78.68           \\ 
                                                                                          & TF-IDF+DT       & 89.09              & 71.70           & 79.46             & 83.43        & 82.60           \\ 
                                                                                          & TF-IDF+XGBoost  & 98.20              & 53.17           & 68.99             & 76.31        & 72.72           \\ 
                                                                                          & TF-IDF+SVM-LR   & 96.35              & 64.39           & 77.19             & 81.52        & 79.70           \\ 
                                                                                          & GloVe+RF        & 91.53              & 26.34           & 40.91             & 62.49        & 50.97           \\ 
                                                                                          & GloVe+DT        & 51.72              & 43.90           & 47.49             & 60.66        & 58.30           \\ 
                                                                                          & GloVe+XGBoost   & 76.92              & 39.02           & 51.78             & 66.28        & 60.42           \\ 
                                                                                          & GloVe+SVM-LR    & 72.17              & 40.48           & 51.88             & 65.94        & 60.83           \\ 
                                                                                          & DeepM1          &  90.16              & \cellcolor{whitesmoke} 80.48           & \cellcolor{whitesmoke} 85.05             & \cellcolor{whitesmoke} 87.82        & \cellcolor{whitesmoke} 87.51           \\ 
                                                                                          & DeepM2          & 86.70              & 79.51           & 82.95             & 86.39        & 86.12           \\ 
                                                                                          & DeepM3          & 90.64              & 61.46           & 73.25             & 78.98        & 77.01           \\  \hline
                                                                                          \multicolumn{2}{c}{Average}           &   87.55            &    58.11        &    68.80        &      76.62   &        73.69   \\  \hline 
\end{tabular}
}}
\end{table*}

\subsubsection{\textbf{Generalizability of Learning Models}}\label{secgeneralization}

\textbf{Motivation}. All learning models are expected to be effective when performing on the new data that they were not trained on, particularly on the data that come from a range of domains (i.e., unseen data) \cite{bousquet2003introduction}, \cite{mlGeneralization}, \cite{bishop2006pattern}. This ability is referred to as generalization. {We aim to measure how well the fifteen learning models detect security discussions in two independent datasets created from different contexts (i.e., out-of-sample accuracy \cite{viviani2019locating}, \cite{aniche2020effectiveness}).}

\textbf{Approach}. First, we created two datasets. The first dataset (i.e., GitHubGen dataset) consists of 500 paragraphs from 2 projects (i.e., \textit{scalecube-services and moleculer} projects) listed in Table \ref{tab:microservices projects}. 250 paragraphs were randomly selected from each project. The second dataset (i.e., StackExGen dataset) is composed of 77 paragraphs (i.e., 77 paragraphs with HTML tag $<$p$>$ from 7 posts) taken from Security Stack Exchange posts with the \say{microservices} tag. The next step included inviting two annotators to determine the presence of security discussions in these 577 paragraphs. Two of the authors acted as an annotator. An author was assigned to annotate 500 paragraphs of GitHubGen dataset, and another author annotated 77 paragraphs of StackExGen dataset. 
After finishing the annotation process, the second author randomly checked 185 of the annotated paragraphs (i.e., confidence level: 90\% and margin of error: 5\%), and any conflicts and disagreements among the annotators were resolved through internal discussions.
It was found that GitHubGen dataset includes 142 security discussions and 358 non-security discussions, and StackExGen dataset consists of 63 security points and 14 non-security discussions. 
{In the last step, we evaluated the previously trained models listed in Table \ref{mldlresults} on GitHubGen dataset, StackExGen dataset, and their combination (i.e., GitHubGen + StackExGen dataset).}



\begin{table*}[h]
\centering
\caption{Two examples of developer discussions labeled by DeepM1 }\label{secclassifiedsample}
\begin{tabular}{|m{12cm}|p{4cm}|}
\hline 
\multicolumn{1}{|l}{\textbf{Security Discussion Samples} }                                                                                                                                                               & \textbf{How DeepM1 predicts?} \\ \hline \hline
\say{@DeividasJackus I agree that microsystems made of components should be isolated and internal calls should not leak outside its boudaries. The issue is that you are thinking of a microsystems as a single node for the sake of optimization and visibility. I think that @icebob already solved the former by calling local service first, and the latter should be resolved with a location independent solution.} [\href{https://github.com/moleculerjs/moleculer/issues/124}{{Taken from issue 124}}\footref{issue124}] & Correctly                     \\ \hline
\say{A general solution I've used in the past is Queue TTL policy. It has its own downsides but it felt a little safer to me.} [\href{https://github.com/moleculerjs/moleculer/pull/341}{{Taken from issue 341}}\footref{issue341}]                                                                                                                                                                                                                                                                                             & Incorrectly                   \\ \hline 
\end{tabular}
\end{table*}

\textbf{Result}. {Table \ref{generalization} shows that the performance of Decision Tree with TF-IDF, DeepM1, and DeepM2 are not significantly degraded across GitHubGen dataset, StackExGen dataset, and GitHubGen + StackExGen dataset. }
{All metrics' values obtained for StackExGen dataset are higher than those for GitHubGen dataset.}
{Particularly, XGBoost With BoW achieves a value of 100\% in the precision metric on GitHubGen dataset. As shown in Table \ref{generalization}, F1-score values obtained from four machine learning algorithms (i.e., RF, DT, XGBoost, and SVM-LR) with GloVe have the worst performance in all datasets. This may stem from the fact that GloVe uses a pre-trained dataset for feature selection.}

\begin{figure}[h]
	\centering
		\includegraphics[scale=.5]{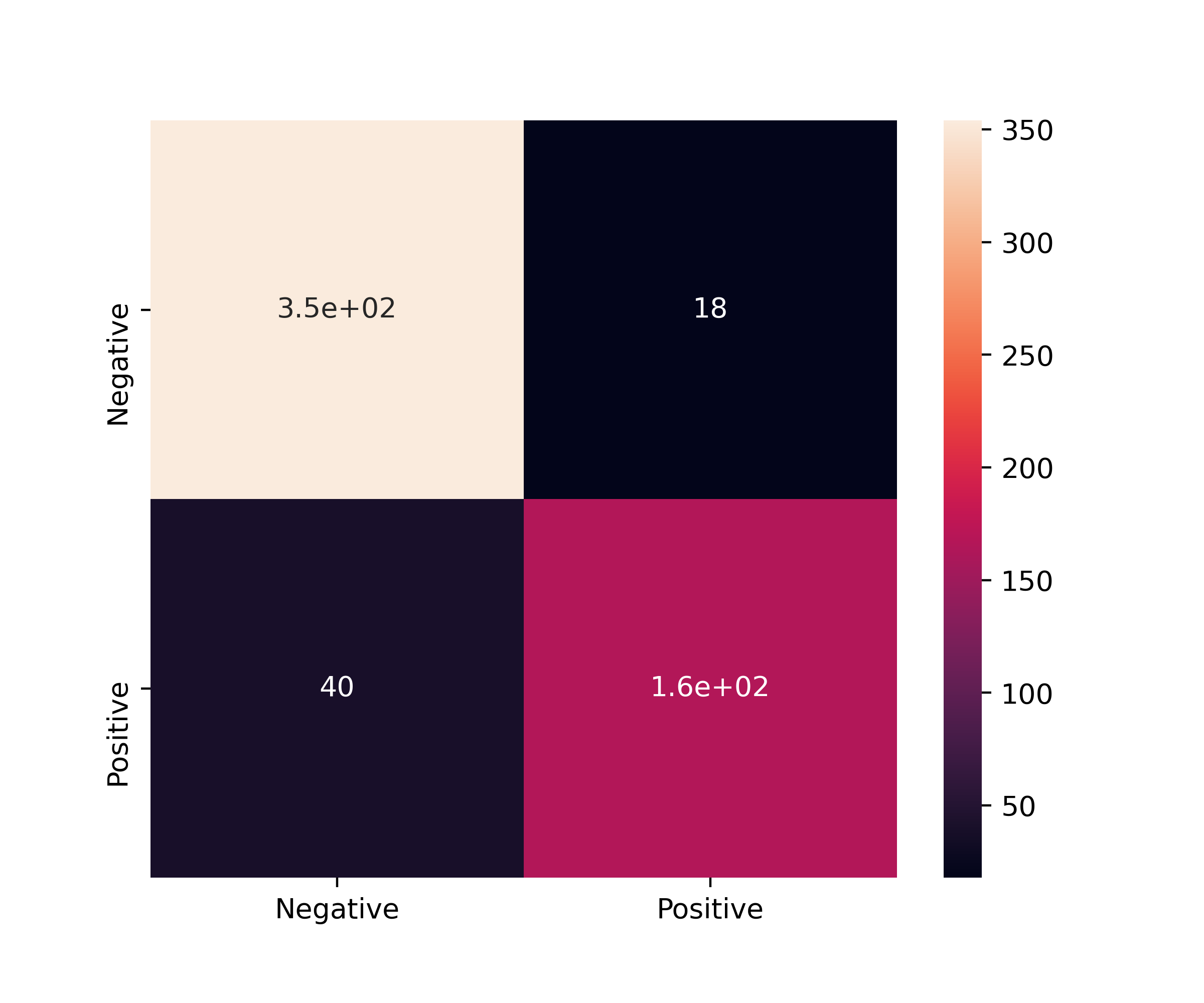}
	\caption{Confusion Matrix of DeepM1 for combination of GitHubGen dataset and StackExGen dataset}
	\label{FIG:seaborn}
\end{figure}

Figure \ref{FIG:seaborn} shows the results of applying DeepM1 on the combination of GitHubGen and StackExGen. DeepM1 could correctly label 519 of 577 paragraphs. Of 58 mislabeled paragraphs, 40 had to be recognized as security discussions, but they were wrongly labeled as a non-security discussion. Table \ref{secclassifiedsample} shows a paragraph from issue 124\footnote{\href{https://github.com/moleculerjs/moleculer/issues/124}{\url{https://bit.ly/3pgxnjg}\label{issue124}}} of the \textit{moleculer} project that was correctly predicted by DeepM1 as a security discussion. It also displays a paragraph from issue 341\footnote{\href{https://github.com/moleculerjs/moleculer/pull/341}{\url{https://bit.ly/3igLEuL}\label{issue341}}} labeled by a human expert as a security discussion, which DeepM1 incorrectly classified it as a non-security discussion.

\begin{table*}[h]
\centering
\caption{Comparison of our best-performing model (DeepM1) with three baselines on unseen datasets. Note that PUMiner\raise0.5ex\hbox{+} and SecureReqNet\raise0.5ex\hbox{+} are a modified version of PUMnier and SecureReqNet, respectively, in which their data pre-processing is replaced with our data pre-processing.}\label{baseline}
\resizebox{15cm}{!}{\begin{tabular}{clc
>{\columncolor{ghostwhite}}c c
>{\columncolor{ghostwhite}}c c
>{\columncolor{ghostwhite}}c }
\hline
\textbf{Models}                                                                           & \textbf{Metrics} & \textbf{GitHubGen} & \textbf{Improv.} & \textbf{StackExGen} & \textbf{Improv.} & \textbf{GitHubGen + StackExGen} & \textbf{Improv.} \\ \hline
                                                                                          & Precision        & 86.71              & ---              & 98.18               & ---              & 90.16                           & ---              \\
                                                                                          & Recall           & 78.16              & ---              & 85.71               & ---              & 80.48                           & ---              \\
                                                                                          & F1-score         & 82.22              & ---              & 91.51               & ---              & 85.05                           & ---              \\
                                                                                          & AUC              & 86.71              & ---              & 89.28               & ---              & 87.82                           & ---              \\
\multirow{-5}{*}{DeepM1}                                                                  & G-mean           & 86.28              & ---              & 89.21               & ---              & 87.51                           & ---              \\ \hline
                                                                                          & Precision        & 23.08              & 3.756 x          & 82.93               & 1.183 x          & 36.41                           & 2.476 x          \\
                                                                                          & Recall           & 23.24              & 3.363 x          & 53.97               & 1.588 x          & 32.68                           & 2.462 x          \\
                                                                                          & F1-score         & 23.16              & 3.550 x          & 65.38               & 1.399 x          & 34.45                           & 2.468 x          \\
                                                                                          & AUC              & 46.26              & 1.874 x          & 51.98               & 1.717 x          & 50.62                           & 1.734 x          \\
\multirow{-5}{*}{Keyword-based Search}                                                               & G-mean           & 40.12              & 2.150 x          & 51.95               & 1.717 x          & 47.33                           & 1.848 x          \\ \hline
                                                                                          & Precision        & 26.01              & 3.333 x          & 89.13               & 1.101 x          & 32.58                           & 2.767 x          \\
                                                                                          & Recall           & 72.53              & 1.077 x          & 65.10               & 1.316 x          & 70.24                           & 1.145 x          \\
                                                                                          & F1-score         & 38.29              & 2.147 x          & 75.23               & 1.216 x          & 44.51                           & 1.910 x          \\
                                                                                          & AUC              & 45.35              & 1.912 x          & 64.68               & 1.380 x          & 45.07                           & 1.948 x          \\
\multirow{-5}{*}{PUMiner}                                                               & G-mean           & 36.29              & 2.377 x          & 64.68               & 1.379 x          & 37.38                           & 2.341 x          \\ \hline
                                                                                          & Precision        & 65.81              & 1.317 x          & 93.88               & 1.045 x          & 74.10                           & 1.216 x          \\
                                                                                          & Recall           & 54.22              & 1.441 x          & 73.02               & 1.173 x          & 60.00                           & 1.341 x          \\
                                                                                          & F1-score         & 59.46              & 1.382 x          & 82.14               & 1.114 x          & 66.31                           & 1.282 x          \\
                                                                                          & AUC              & 71.53              & 1.212 x          & 75.79               & 1.177 x          & 74.22                           & 1.183 x          \\
\multirow{-5}{*}{SecureReqNet}                                                               & G-mean           & 69.40              & 1.243 x          & 75.74               & 1.177 x          & 72.84                           & 1.201 x          \\ \hline
                                                                                          & Precision        & 44.94              & 1.929 x          & 82.81               & 1.185 x          & 54.96                           & 1.640 x          \\
                                                                                          & Recall           & 56.34              & 1.387 x          & 84.12               & 1.018 x          & 64.88                           & 1.240 x          \\
                                                                                          & F1-score         & 50.00              & 1.644 x          & 83.46               & 1.096 x          & 59.51                           & 1.429 x          \\
                                                                                          & AUC              & 64.48              & 1.344 x          & 52.77               & 1.691 x          & 67.79                           & 1.295 x          \\
\multirow{-5}{*}{\begin{tabular}[c]{@{}c@{}}PUMiner\raise0.5ex\hbox{+}\end{tabular}} & G-mean           & 63.96              & 1.348 x          & 42.46               & 2.101 x          & 67.73                           & 1.292 x          \\ \hline
                                                                                          & Precision        & 64.41              & 1.364 x          & 94.23               & 1.041 x          & 73.53                           & 1.226 x          \\
                                                                                          & Recall           & 53.52              & 1.460 x          & 77.78               & 1.101 x          & 60.98                           & 1.319 x          \\
                                                                                          & F1-score         & 58.46              & 1.406 x          & 85.22               & 1.073 x          & 66.67                           & 1.275 x          \\
                                                                                          & AUC              & 70.89              & 1.223 x          & 78.17               & 1.142 x          & 74.44                           & 1.179 x          \\
\multirow{-5}{*}{\begin{tabular}[c]{@{}c@{}}SecureReqNet\raise0.5ex\hbox{+}\end{tabular}} & G-mean           & 68.73              & 1.255 x          & 78.17               & 1.141 x          & 73.21                           & 1.195 x          \\ \hline
\end{tabular}}
\end{table*}

\subsubsection{Comparison with Baselines}\label{sec:CompBaseLines}
\textbf{Motivation.} 
To the best of our knowledge, no study develops learning models to identify security discussions from microservices practitioner discussions. Despite this, we compared our best-performing model (DeepM1) with three baselines. Similar to \cite{alomar2021finding}, \cite{obie2020first}, \cite{sorbo}, we developed a keyword-based search as the first baseline. The result of the keyword-based search also helps us understand if the identification of security discussions in microservices systems is a learning problem or not \cite{alomar2021finding}. Two studies \cite{le2020puminer} and \cite{palacio2019learning} are the closest works to our study. The study \cite{le2020puminer} developed PUMiner to identify security posts in Stack Overflow, and the study \cite{palacio2019learning} introduced SecureReqNet to distinguish security issues from non-security issues in the GitHub repository. We used these two approaches as two other baselines.

\textbf{Approach.} Here we describe how the three baselines are implemented.

\underline {Keyword-based Search (Baseline1)}. As we described in Section \ref{secBuilddataset}, the annotators were asked to identify security-related words in security discussions. We used these security-related words to build our keyword-based search. Some of these security-related words appeared in security discussions as abbreviations, and others were in full forms. For example, we had \say{single sign on}, \say{sso}, and \say{single-sign-on} in the collected security discussions. In the first step, we converted such security-related words to one format (i.e., abbreviation). This resulted in 165 unique security keywords. We then applied the stemming process (see Step 4 in Section \ref{secPreprocess}) on paragraphs in unseen datasets. Like \cite{obie2020first}, \cite{sorbo}, our keyword-based search calculates the probability of a stemmed paragraph to include one or more of the 165 unique security keywords. To be more precise, suppose \textit{\textbf{P}} is a stemmed paragraph, and \textit{\textbf{SL}} is a list including all 165 unique security keywords. We define \textit{\textbf{N\textsubscript{SL}}} as the number of tokens in \textbf{\textit{P}} that emerge in \textit{\textbf{SL}} and \textbf{\textit{N\textsubscript{P}}} as the number of tokens in \textbf{\textit{P}}. The keyword-based search measures the probability that \textit{\textbf{P}} is classified as a security discussion with the following formula: \textbf{\textit{$Pr\textsubscript{(P,SW)} = \frac{N\textsubscript{SW}}{N\textsubscript{P}}$}}. It should be noted that there might be cases where some paragraphs are mistakenly labeled as security discussion because just one security-related word is included in \textbf{\textit{P}}. To avoid this, the keyword-based search labels \textbf{\textit{P}} as a security discussion if \textbf{\textit{$Pr\textsubscript{(P, SW)}$}} is higher than 0.05 (i.e., security words compose at least 5\% of a paragraph) \cite{obie2020first}, \cite{sorbo}. 

\underline{PUMiner (Baseline2) and SecureReqNet (Baseline3)}.
To make the comparison fair, we replicated PUMiner and SecureReqNet in two distinct ways. First, we replicated them with their own data pre-processing and trained them with our dataset using 10-fold cross-validation. Second, we replaced PUMiner's and SecureReqNet's data pre-processing with our data pre-processing described in Section \ref{secPreprocess} and call them PUMiner\raise0.5ex\hbox{+} and SecureReqNet\raise0.5ex\hbox{+} respectively. We trained PUMiner\raise0.5ex\hbox{+} and SecureReqNet\raise0.5ex\hbox{+} with our dataset using 10-fold cross-validation. We evaluated the performance of all baselines (the keyword-based search and the trained PUMiner, SecureReqNet, PUMiner\raise0.5ex\hbox{+}, SecureReqNet\raise0.5ex\hbox{+}) on unseen datasets (GitHubGen, StackExGen, and their combination) described in Section \ref{secgeneralization}.

{\textbf{Result.} 
Table \ref{baseline} shows that DeepM1 outperforms all baselines on all unseen datasets in all metrics, with improvements ranging from 1.018x to 3.756x. SecureReqNet\raise0.5ex\hbox{+} has only comparable performance with DeepM1 on StackExGen dataset in the recall metric (i.e., recall value achieved by SecureReqNet\raise0.5ex\hbox{+} is 84.12\% versus 85.71\% recall value obtained by DeepM1). Except for this, the performance of all baselines is significantly lower than that of DeepM1 in all metrics on unseen datasets.} Also, as shown in Table \ref{baseline}, PUMiner\raise0.5ex\hbox{+} and SecureReqNet\raise0.5ex\hbox{+} have a better performance than when they use their own pre-processing (i.e., PUMiner and SecureReqNet).

\begin{center}
\begin{tcolorbox}[colback=gray!5!white,colframe=black!75!black]
\textbf{RQ2 Summary.} \textit{The results show that the performances of both ML and DL models in identifying security discussions are promising: \textbf{\textbf{precision ($\overline{84.86}$\%)}, recall ($\overline{72.80}$\%)}, \textbf{F1-score ($\overline{77.89}$\%)}, \textbf{AUC ($\overline{83.75}$\%)}, and \textbf{G-mean ($\overline{82.77}$)}. However, DeepM1, as a deep learning model, is the best performing learning model. Additionally, DeepM1's performance of detecting security discussions on unseen datasets (generalizability) is not significantly degraded. Finally, the evaluation demonstrates that DeepM1 outperforms the state-of-the-art baselines with improvements ranging from 1.018x to 3.756x in all metrics.}
\end{tcolorbox}
\end{center}

\begin{figure*}[t]
	\centering
		\includegraphics[scale=.70]{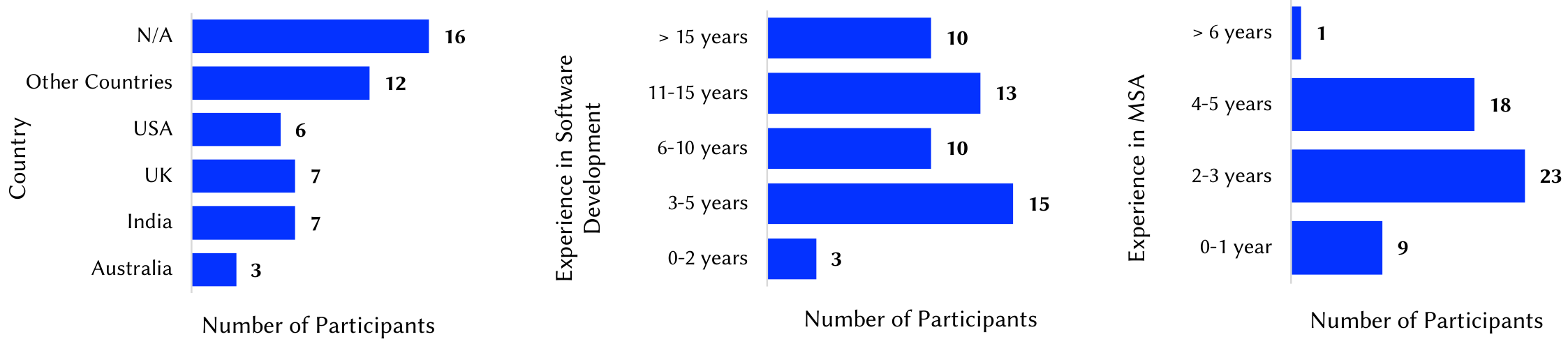}
	\caption{The validation survey: participants demographic data (n=51)}
	\label{FIG:demoSurvey2Result}
\end{figure*}

\begin{figure*}[t]
	\centering
		\includegraphics[scale=.65]{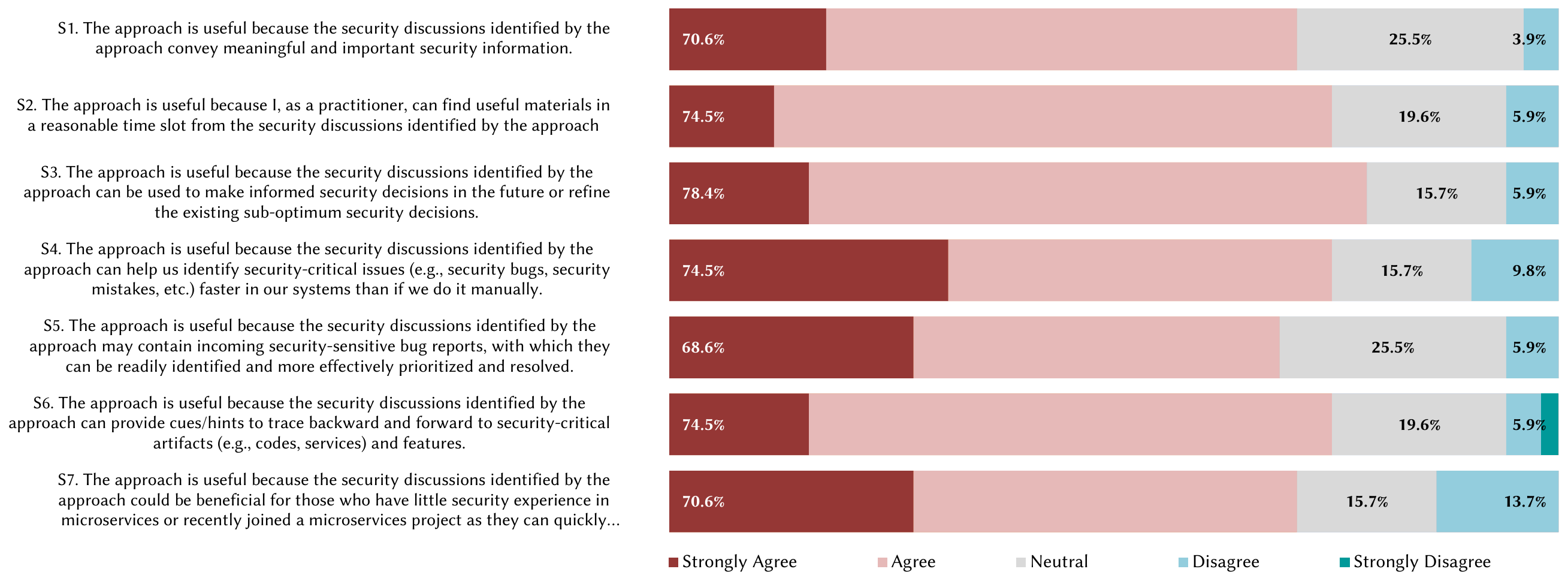}
	\caption{The validation survey: responses to the statements regarding the usefulness of detected security discussions using the best learning model (n=51)}
	\label{FIG:likerResult}
\end{figure*}

\subsection{\textbf{Usefulness of Automated Identification of Security Discussions (RQ3)}} \label{secvalidationresulst}

As presented in Section \ref{validationsurvey}, we conducted a survey (i.e., validation survey) to solicit the perceptions of practitioners on the usefulness of the security discussions detected by the best-performing learning model (i.e., DeepM1).

{Figure \ref{FIG:demoSurvey2Result} shows the demographic information of 51 practitioners who completed the validation survey. 16 participants did not reveal which country they were working in. The rest 35 respondents came from 16 countries, in which India and UK, each with seven participants, were dominant. Almost 65\% (33) of the respondents indicated that they had more than six years of software development experience in the industry. 15 participants had 3-5 years of experience, and three respondents developed software for less than two years. Of the 51 respondents, nine had less than one year of experience in microservices system development, 23 had 2-3 years experience, 18 had 4-5 years experience, and one had more than six years experience.}

{Figure \ref{FIG:likerResult} shows the feedback of the 51 practitioners on the usefulness of the results (e.g., distinguished security discussions from non-security discussions) generated by the best-performing learning model (i.e., DeepM1). We observed that almost 71\% of the respondents (strongly) agreed that the security discussions identified by DeepM1 convey meaningful and important security information (Statement S1). 25.5\% of the participants were neutral about Statement S1, and only 3.9\% disagreed. They further indicated that identifying useful materials and information from the detected security discussions can be done in a reasonable time slot (Statement S2). 74.5\% rated this statement as \textit{strongly agree} or \textit{agree}.}

{Statements S3 to S7 investigate how the detected security discussions can be helpful. Approximately 79\% of the respondents answered Statement S3 as \textit{strongly agree} or \textit{agree}, indicating that security discussions help them make informed security decisions in the future or refine the existing sub-optimum security decisions. Another benefit that the security discussions extracted from developer discussions can bring is to help practitioners find security-critical issues (e.g., security bugs, security mistakes) in their systems faster than if they do it manually (Statement S4). 74.5\% strongly agreed or agreed with Statement S4, and 9.8\% disagreed with this statement. In close to 70\% of the responses to Statement S5, the respondents felt that the identified security discussions could help trace backward and forward to security-critical artifacts (e.g., codes, services) and features. Another benefit of security discussion is that security-sensitive bug reports can be readily identified and more effectively prioritized and resolved. It is because security discussions may contain incoming security-sensitive bug reports (74.5\% strongly agreed or agreed with Statement S6, 19.6\% were neutral, and 5.9\% strongly disagreed or disagreed with this statement). Regarding Statement S7, 70.6\% believed that the information in the identified security discussions could help newcomers and less experienced team members quickly access and learn security solutions and avoid common security mistakes.}

\begin{center}
\begin{tcolorbox}[colback=gray!5!white,colframe=black!75!black]
\textbf{RQ3 Summary:} \textit{{The validation survey respondents confirm that the security discussions detected by the best-performing learning model (i.e., DeepM1) have promise for application into securing microservices systems, particularly in helping practitioners make informed security decisions in the future or refine the existing sub-optimum security decisions.}}
\end{tcolorbox}
\end{center}

\section{Discussion}\label{secDiscuss}

{This section discusses the implications for practitioners and researchers based on our key findings and analysis of the responses to an open-ended question: \say{\textit{What improvements to our approach would further help develop secure microservices systems?}}.}

{
\textbf{Better support is still needed}. The challenging nature of security in the MSA style (confirmed by the preliminary survey) poses difficulties for all microservices practitioners in general, and novice and non-security microservices developers in particular, to comprehend security in microservices systems and effectively adopt and implement best security practices. The validation survey confirms that microservices practitioners can identify and learn succinct security knowledge with reasonable time and effort from the results (security discussions) produced by the learning models.  This matter becomes much more critical when we know that the security discussions investigated in this study include security solutions and principles. This could be more beneficial for those who have little security experience or recently joined a microservices project. They can quickly be aware of the up-to-date security solutions and best practices to secure microservices systems. However, our participants emphasized that microservices practitioners still need more support to design and implement secure microservices systems. New tools can be developed, or our models can be extended to automatically link security discussions in developer discussions to relevant learning documents and materials (e.g., blogs, code snippets). Further to this, for a detected security issue in a given microservices system, new tools should be able to come up with possible security solutions with relevant references.}

{
\textbf{Prioritize security discussions}. Our learning models can only distinguish security discussions from non-security discussions. Our validation survey reveals that security discussions detected from developer discussions can help developers to identify security-critical services or security-critical issues (e.g., security bugs) in microservices systems. The importance of identifying such security-critical services and issues in the MSA style is increasing as the entire system may go down due to a single compromised microservice \cite{sun2015security}, \cite{yarygina2018overcoming} . Despite this, the participants commented that security discussions could be labeled with different priority levels, such as critical, important, or low-impact. Determining the types and severity of security discussions can help a software team assign expert members to address and audit more security-critical issues in a faster time \cite{palacio2019learning}. Hence, there is a need to develop tools that label security discussions based on their types and severity.}

{\textbf{Design context and security discussions.} In this study, we chose to label and identify security discussions at the level of the paragraph. Viviani et al. \cite{viviani2019locating} also identified design points (\say{\textit{a piece of a discussion relating to a decision about a software system’s design}}) at the level of the paragraph in developer discussions. This decision was made to enable microservices practitioners to capture, identify, and share meaningful and succinct security knowledge faster and more accurately without going through an entire developer discussion to find and understand a security issue, decision, or solution. Despite this, developers may need more design context information to thoroughly understand some security discussions detected by our learning models and effectively apply security solutions discussed in the detected security discussions in practice. The participants mentioned the need to develop tools that can extract the relevant design context information for each detected security discussion and make them available as extra and optional resources (e.g., through external links) for microservices practitioners. Such design context information can even be extracted from non-security discussions.}

\section{Threats to Validity}\label{secthreats}
This section lists possible threats to validity related to our research method and findings and the adopted mitigating strategies to these threats. We classify the threats into three types: internal, construct, and external \cite{wohlin2012experimentation}. 

\subsection{\textbf{Internal Validity}}\label{secIntenalval}

\textbf{Surveys}. {The selection of microservices practitioners was a challenge for our surveys. In the preliminary survey, the participants were either the contributors to open-source microservices systems or the contributors to open-source projects that develop tools and frameworks for microservices systems. We also invited microservices practitioners working in the industry for the validation survey after analyzing their LinkedIn profiles. We are confident that most participants had enough experience and expertise in the MSA style. Specific roles may produce bias in the survey results. While software engineers, developers, and architects were the dominant roles in our surveys, other roles such as DevOps engineers, team leads, testers, etc., participated in our survey as well.}

\textbf{Experiments}. The annotation process may have introduced two threats. The first one is the annotators' subjective bias. We implemented two strategies to minimize this threat. (1) We carried out a pilot study to reach a common understanding of the characteristics of security and non-security discussions. This step also helped us develop a robust coding schema for the annotation process. (2) We employed several annotators in the pilot annotation process and the main annotation process to reduce the potential personal bias (i.e., investigator triangulation \cite{carter2014use}). The second threat that might have happened is that the annotators erroneously labeled the paragraphs. The measure taken to minimize this threat was that the second author, who was not involved in the annotation process, cross-checked 581 paragraphs out of 17,277 paragraphs (i.e., confidence level: 95\% and margin of error: 4\%). We also obtained the value of 0.82 for Cohen’s Kappa Coefficient over the 581 cross-checked paragraphs, indicating an agreement between the cross-checker and the annotators.
The second author held several meetings with the annotators to discuss the annotations and resolve any disagreements. Furthermore, the annotators had extensive experience in coding, design, and microservices systems. They ranged from a software engineer to a professor. With these efforts, we believe that the dataset is credible with minimum mislabeled paragraphs.

Given our unit of analysis is the paragraph, as we discussed in Section \ref{secDiscuss}, our learning models may identify security discussions that lack enough design context information. We accept that practitioners may not understand completely some security discussions detected by our learning models. However, we believe that most security discussions (e.g., SD1 to SD4 in Figure \ref{FIG:motivatingscenario}) identified by our learning models include useful and necessary security information.

\subsection{Construct Validity}\label{secConstractval}

\textbf{Surveys}. Practitioners may have different interpretations and understandings of \say{security discussion} in microservices systems. {Our strategy to moderate this threat was defining \say{security discussion} at the beginning of both the preliminary survey and validation survey}. We also provided some examples of security discussions from GitHub and Stack Overflow. It should be noted that none of the participants showed disagreement with our definition regarding security discussions. We only used one research method (the validation survey) to evaluate the usefulness of our approach (i.e., RQ3). Our validation survey may not have shown precisely and comprehensively all benefits and possible deficiencies (e.g., creating additional effort for practitioners) of our approach in the daily work of microservices practitioners. Other research methods, such as controlled experiments and in-depth qualitative studies, should be conducted to assess different aspects of our approach.

\textbf{Experiments}. The selection of metrics and ML/DL models can be another source of threat. We used a wide variety of metrics to show various aspects of the performance of the learning models.
Among these four metrics, recall, precision, and F1-score metrics are widely used and highly recommended in the software engineering literature (e.g., \cite{viviani2019locating}, \cite{li2020ontology}, \cite{abualhaija2019machine}, \cite{lessmann2008benchmarking}, \cite{romano2011using}). As our dataset was imbalance, we employed the G-mean metric to show the effectiveness of ML/DL models with one number \cite{goseva2018identification}. This work used fifteen classification models, including ML and DL models, among many possible learning models. Peters et al. \cite{peters2017text} argued that it is not possible to use all classifiers in a single case study. It was also found that out of the 22 studied ML classifiers, the top 17 classifiers do not significantly differ in terms of performance \cite{lessmann2008benchmarking}. Three ML classifiers used in our study are from the 17 top-ranked classifiers. 
We also built three DL models using Bi-LSTM and CNN. The DL models based on Bi-LSTM and CNN have shown desired results \cite{zhao2017learning}. Thus, we acknowledge that our experiment results may differ if we use other ML/DL models.

\subsection{\textbf{External Validity}}\label{secExternalval}
\textbf{Surveys}. {The number of responses gathered from the preliminary survey and validation survey can limit their findings. We only received 67 responses for the preliminary survey and 51 responses for the validation survey. Further, the majority of the participants who completed the preliminary survey were contributors to open-source projects. Hence, most observations in the preliminary survey are exclusive to open-source microservices systems. In the validation survey, we tried to reduce this threat by recruiting practitioners who worked in the software industry and contributors to open-source projects. However, we acknowledge that the results obtained from the surveys may not be generalized to all microservices practitioners, types of projects (e.g., closed source software), and companies.}

\textbf{Experiments}. 
The dataset used in this work is composed of 17,277 paragraphs (i.e., 4,813 security discussions and 12,464
non-security discussions) extracted from two sources: issue discussions collected from five relatively large-scale open-source microservices systems and 498 Stack Overflow posts. We also applied all developed learning models to 500 unseen paragraphs from two open-source microservices systems and 77 unseen paragraphs from seven Security Stack Exchange posts with the \say{microservices} tag (Section \ref{secgeneralization}). Hence, we admit that these open-source projects and Stack Overflow posts are not representative of all projects hosted on GitHub and Stack Overflow posts. Another limitation is that we only focused on issue discussions in the selected projects. Our dataset could be more comprehensive by considering and analyzing the other elements of the selected open-source projects, such as their source codes, comments, commits, and design documents. Finally, we acknowledge that our findings may not be generalizable, particularly to projects from different domains, closed source projects, and projects hosted on other platforms such as GitLab and BitBucket. We have made our datasets and the implementation of ML and DL models publicly available \cite{onlinedataset} to enable other researchers and practitioners to use, replicate, validate, or extend our study.
 
\section{Related Work}\label{secRelatework}
To the best of our knowledge, there is no study that uses learning models to identify security information in developer discussions (such as {GitHub} issues and Stack Overflow posts) of microservices systems. We summarize the studies that discuss security support in microservices systems (Section \ref{secSecuMSA}), and discuss the studies that use ML- and DL-based models to recognize security contents in textual software artifacts (Section \ref{secMLforSec}). In Section \ref{secDifWithOthers}, we discuss the differences of our study with existing studies.

\subsection{\textbf{Security in Microservices Systems}}\label{secSecuMSA}

Pereira et al. \cite{Pereira2019SecMec} conducted a systematic mapping of security mechanisms in MSA, revealing that most studies focus on \say{unauthorized access}, \say{sensitive data exposure}, and \say{compromising individual microservices}. They also found that prevention and auditing based solutions are widely used to secure microservices. In another review study, Waseem et al. \cite{waseem2020systematic} observed a lack of concrete solutions to achieve and address monitoring, security, and performance in microservices systems in DevOps.
Yu et al. \cite{yu2019survey} surveyed and classified 19 security issues related to microservices communication into four categories: containers (e.g., kernel exploit, malicious process), data (e.g., data intercept, secret leak), permission (e.g., identify spoofing), and network (SDN issues, DOS attack). Based on identified issues, they proposed a solution to address security issues in microservices-based fog applications.
Ghofrani and Lübke \cite{ghofrani2018challenges} surveyed 25 microservices experts to evaluate current practices and challenges in the MSA style. 
One of the main conclusions of their survey is that practitioners prioritize security, response time, and performance over other quality attributes (e.g., resilience) in microservices systems.

Sun et al. \cite{P2} built a monitoring and policy enforcement infrastructure using API primitive FlowTap. Their solution provides security-as-a-service to detect and block internal and external threats over the network for microservices-based cloud systems. Pahl and Donini \cite{sun2015security} introduced a certificate-based (X.509) method for securing IoT microservices. The method helps the distributed nodes verify the security properties locally and change certificate properties across the distributed IoT nodes. A multilateral (e.g., confidentiality, accountability, and integrity) security framework was proposed in \cite{P3} to assess the security design and architecture quality for docker-based applications (e.g., microservices). The security framework provides multi-layered security by considering the concepts of TCP/IP, OSI stack, and cloud stack model. Ahmadvand and Ibrahim \cite{P4} suggested a method that considers security and scalability as the first-class citizens when decomposing monolithic systems into microservices. Their approach uses an asset-threat-risk triangle to elicit and capture security requirements and decisions.

The increasing importance of security in microservices systems and the lack of concrete security solutions and guidelines for such systems inspired us to explore and extract the security knowledge scattered in practitioner discussions.

\subsection{\textbf{Learning Models for Recognizing Security Contents}}\label{secMLforSec}

Yang et al. \cite{yang2016security} identified 30 security topics (e.g., SQL injection, Timing attack) on Stack Overflow posts using Latent Dirichlet Allocation (LDA) tuned by the Genetic Algorithm (GA). The 30 topics are grouped into \say{web security}, \say{mobile security}, \say{system security}, \say{cryptography}, and \say{software security}. 
Le et al. \cite{le2020puminer} introduced PUminer to identify security posts from {Stack Overflow} and Security Stack Exchange posts using a small number of labeled security posts when a considerable amount of posts are unlabeled. Based on the experiments on more than 17.3 million posts, PUMiner showed more than 0.85\% F1-score and G-mean in recognizing security posts from {Stack Overflow} and Security Stack Exchange.

Palancio et al. \cite{palacio2019learning} used several variations of SecureReqNet’s CNN architecture to distinguish security issues from non-security ones in GitHub and GitLab. The SecureReqNet approach yielded, on average, an accuracy of 81.29\% on open-source issues and 69.77\% on the requirements of commercial software systems. Han et al. \cite{han2017learning} developed a DL approach to determine the difficulty level of a software vulnerability based on the vulnerability description. The approach's assessment showed its effectiveness in short-text vulnerability descriptions.

Pletea et al. \cite{pletea2014security} constructed a dataset of security-related terms to carry out sentimental analysis of security discussions on GitHub. Marrison et al. \cite{morrison2018identifying} found that each project or domain has unique security keywords, and using project-specific security keywords for predicting security issues of the same project leads to better performance compared to using general security keywords or security keywords from other projects. 
Tao et al. \cite{tao2020identifying} introduced a Security-Related Review Miner (SRR-miner) that uses a keyword-based technique to extract security-related review sentences on 17 mobile applications in Google Play.

A set of studies have developed ML-based approaches to discriminate security requirements from other types of requirements (e.g., \cite{bettaieb2019decision}, \cite{li2020ontology}, \cite{abualhaija2019machine}, \cite{knauss2011supporting}). In \cite{bettaieb2019decision}, the focus was on identifying security controls, an important task in assessing security requirements, and it was found that applying cost-sensitive learning (CSL) and SMOTE (Synthetic Minority Over-sampling TEchnique) methods to imbalanced classes would improve the performance of classifiers.

Others have focused on identifying security bug reports (SBRs). Peters and colleagues \cite{peters2017text} proposed a framework to decrease the chance of mislabeling SBRs. The proposed framework achieved this goal by leveraging filtering and ranking techniques. 
In another study, Jiang et al. \cite{jiang2020ltrwes} utilized  \textit{learning to rank} and \textit{word embedding} techniques for this purpose.
While the previous works use the entire bug reports, Pereira et al. \cite{pereira2019identifying} predicted SBRs only based on the title of SBRs and in the presence of noisy labels (0.98\% AUC).

\subsection{Differences with Existing Studies}\label{secDifWithOthers}
{While the techniques mentioned above yield acceptable performance, our focus, context, and findings are different from the existing techniques:}

{\textbf{(1)} They are not specifically designed for MSA. It is argued that apart from standard security terms (e.g., password), which may be found in the security texts of any projects or domains, each project or domain has its own unique security-related terms \cite{morrison2018identifying}, \cite{knauss2011supporting}. Our preliminary survey findings also reveal that security concerns in microservices systems are unique compared to other types of systems, such as monoliths. This denotes that security discussions that happen among microservices practitioners are unique.}

{\textbf{(2)} Our work is different from the studies (i.e., \cite{bettaieb2019decision}, \cite{li2020ontology}, \cite{abualhaija2019machine}, \cite{knauss2011supporting}) focusing on distinguishing security requirements from non-security requirements. Requirements specifications mostly focus on functional requirements and lack the security design decisions, design rationale, the security issues faced by developers, and security solutions employed to address security challenges.}

{\textbf{(3)} The techniques proposed in \cite{peters2017text}, \cite{jiang2020ltrwes}, \cite{pereira2019identifying} to discriminate security bug reports from non-security bug reports are also different from our work.
An issue in an issue tracking system might include various information, such as a task, new feature, or bug \cite{bhat2017automatic}. Hence, bug reports can only capture, \textit{if any}, a small part of developer discussions and decisions \cite{goseva2018identification}. Moreover, as a bug report is usually written when a bug happens, it does not document all security issues developers face and the security solution options they consider.}

{\textbf{(4)} One key design decision that we made when developing our dataset and learning models was to focus on paragraphs. In contrast to this key decision, Le et al. \cite{le2020puminer} tried to distinguish security contents from non-security ones at the granularity of an Stack Overflow post (i.e., each post includes a question, a list of answers, and the appended comments \cite{zhang2019reading}). The studies \cite{palacio2019learning}, \cite{morrison2018identifying} do the same task at the granularity of an issue (i.e., with all comments on the issue) in issue tracking systems. We made that decision because Viviani et al. \cite{viviani2019locating} found that a design decision can be presented and identified in a paragraph. Moreover, Xu et al. \cite{xu2017answerbot} revealed that a paragraph could summarize useful information of a lengthy answer post in Stack Overflow.

GitHub issues and Stack Overflow posts, particularly the lengthy ones, may discuss different topics (e.g., security, performance) \cite{viviani2019locating}. It may also be difficult for practitioners to identify useful information from lengthy posts and issues \cite{xu2017answerbot}. Imagine post number 2283937\footnote{\href{https://stackoverflow.com/questions/2283937}{{\url{https://bit.ly/369XnFF}}}} and issue number 107 \footref{eshop107}, which have 25872 and 14233 words, respectively, and discuss different topics. Suppose the English silent reading rate is 238 words per minute \cite{brysbaert2019many}. If post number 2283937 is correctly labeled as security, a developer will still spend one hour and fifty minutes to read this post and probably spend more time to find which part(s) of the post discusses security. Hence, we argue that the security discussions detected by our learning models, compared to the ones identified by the models in \cite{le2020puminer} and \cite{palacio2019learning}, include succinct security knowledge. This enables practitioners to find the required security information from such security discussions in a more reasonable time and with less effort.}

{\textbf{(5)}} Among the learning models mentioned above aiming to discriminate between security contents and non-security contents, PUMiner \cite{le2020puminer} and SecureReqNet \cite{palacio2019learning} are the closest to our work. 
The comparison conducted in Section \ref{sec:CompBaseLines} shows that our best-performing learning model (DeepM1) significantly outperformed two configurations of PUMiner by 42.91-91.08\% and two configurations of SecureReqNet by 27.56-28.26\% in F1-score when they were applied on an unseen dataset including both GitHub and Security Stack Exchange data. Further to this, the performance of PUMiner and SecureReqNet on GitHubGen unseen dataset (GitHub data) was lower than once they were applied on StackExGen unseen dataset (Security Stack Exchange data) and the combination of GitHubGen and StackExGen dataset. Several factors may explain why DeepM1 achieves significantly higher performance. (1) PUMiner and SecureReqNet need an entire Stack Overflow post or GitHub issue to differentiate security contents from non-security contents. At the same time, DeepM1 needs smaller chunks of information (paragraphs) from the GitHub issue or Stack Overflow post for this purpose. (2) As discussed earlier and confirmed by the preliminary survey, the nature of security discussions in microservices systems is different from security discussions in other systems. (3) PUMiner needs an ultra-large-scale data input for training to show its performance. (4) DeepM1 uses a combination of CNN and LSTM networks to consider both different textual features and context information in developer discussions.

\section{\textbf{Conclusion and Future Work}}\label{secConclusion}

There is a knowledge gap among practitioners and organizations in designing and implementing secure microservices systems. This study tried to (partially) bridge this gap by automatically identifying security discussions from previous microservices developer discussions.
We first conducted a preliminary survey with 67 microservices practitioners to understand their perspectives on security in microservices systems. The findings of the preliminary survey showed that securing microservices systems is a unique challenge for software practitioners. The survey respondents also confirmed the usefulness of (automatically) identifying and then leveraging developer security discussions scattered in existing microservices systems in making security decisions.
 
Hence, we have developed fifteen {machine/deep learning} models to identify security discussions from developer discussions ({GitHub} issues and Stack Overflow posts) of microservice systems. Our results show that DeepM1, a deep learning model, performs the best. On average, the developed machine/deep learning models could achieve 84.86\% precision, 72.80\% recall, 77.89\% F1-score, 83.75\% AUC, and 82.77\% G-mean. We further solicited microservices practitioners' perception of the usefulness of the results produced by DeepM1 through another survey (i.e., validation survey). The respondents to the validation survey acknowledged that the security discussions detected by DeepM1 could have promising applications in practice. Notably, the security discussions can help practitioners make informed security decisions in the future or refine the existing sub-optimum security decisions and identify security-critical issues (e.g., security bugs, security mistakes, etc.) faster in microservices systems.

We plan to apply our models to a range of commercial microservices systems in the future. We also intend to qualitatively analyze security discussions to build an evidence-based knowledge about best security practices for microservices systems. {Further to this, we aim to extend our models to identify and classify the types of security decisions in developer discussions.}

\section*{\textbf{Acknowledgements}}

We would like to thank all the practitioners who participated in the surveys. This work is partially sponsored by the National Key R\&D Program of China with Grant No. 2018YFB1402800.

 \bibliographystyle{elsarticle-num} 
 \bibliography{Main}





\end{document}